\documentclass[lettersize,journal]{IEEEtran}
\usepackage{amsmath,amsfonts}
\usepackage{array}
\usepackage[caption=false,font=normalsize,labelfont=sf,textfont=sf]{subfig}
\usepackage{textcomp}
\usepackage{stfloats}
\usepackage{url}
\usepackage{verbatim}
\usepackage{graphicx}

\def\BibTeX{{\rm B\kern-.05em{\sc i\kern-.025em b}\kern-.08em
    T\kern-.1667em\lower.7ex\hbox{E}\kern-.125emX}}
\usepackage{balance}

\usepackage{etoolbox}
\usepackage{enumitem}
\usepackage{subcaption}
\usepackage{multirow}
\usepackage{hhline}
\usepackage{rotating}
\usepackage{algorithm}
\usepackage{algpseudocode}
\usepackage{soul}
\usepackage{xcolor}

\usepackage{booktabs}

\newcommand{\rev}[1]{{\color{black}#1}} %revision

\newcommand{\system}{Synergy}

\begin{document}
\title{\system{}: Towards On-Body AI via Tiny AI Accelerator Collaboration on Wearables}
\author{Taesik Gong, SiYoung Jang, Utku Günay Acer, Fahim Kawsar, and Chulhong Min 
\thanks{Taesik Gong, who is now with UNIST (e-mail: taesik.gong@unist.ac.kr), was with Nokia Bell Labs when this work was done. SiYoung Jang is with Nokia Bell Labs (e-mail: siyoung.jang@nokia-bell-labs.com). Utku Günay Acer is with Nokia Bell Labs (e-mail: utku\_gunay.acer@nokia-bell-labs.com). Fahim Kawsar is with Nokia Bell Labs and University of Glasgow (e-mail: fahim.kawsar@nokia-bell-labs.com). Chulhong Min, the corresponding author, is with Nokia Bell Labs (e-mail: chulhong.min@nokia-bell-labs.com).}}

\markboth{}%
{How to Use the IEEEtran \LaTeX \ Templates}

\maketitle

\begin{abstract}

The advent of tiny artificial intelligence (AI) accelerators enables AI to run at the extreme edge, offering reduced latency, lower power cost, and improved privacy. When integrated into wearable devices, these accelerators open exciting opportunities, allowing various AI apps to run directly on the body. We present \emph{\system{}} that provides AI apps with best-effort performance via system-driven holistic collaboration over AI accelerator-equipped wearables. To achieve this, \system{} provides device-agnostic programming interfaces to AI apps, giving the system visibility and controllability over the app's resource use. Then, \system{} maximizes the inference throughput of concurrent AI models by creating various execution plans for each app considering AI accelerator availability and intelligently selecting the best set of execution plans. \system{} further improves throughput by leveraging parallelization opportunities over multiple computation units. Our evaluations with 7 baselines and 8 models demonstrate that, on average, \system{} achieves a 23.0$\times$ improvement in throughput, while reducing latency by 73.9\% and power consumption by 15.8\%, compared to the baselines.

\end{abstract}


\section{Introduction}~\label{sec:introduction}

\begin{figure}[t!]
\centering
\includegraphics[width=0.8\linewidth]{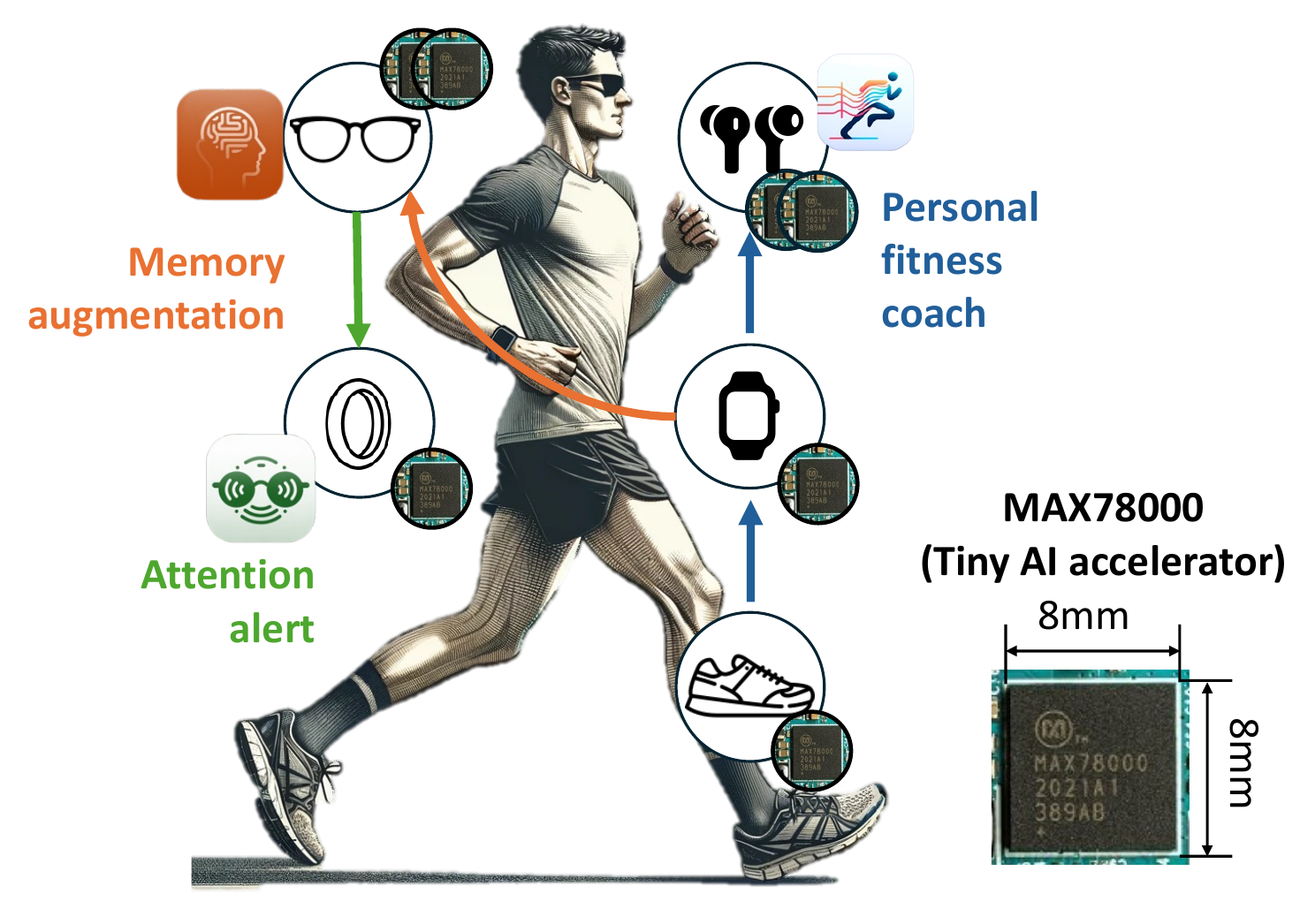}
%\vspace{-.4cm}
\caption{Wearable computing powered by tiny AI accelerators, simultaneously running on-body AI apps.}
\vspace{-.5cm}
\label{fig:motivation}
\end{figure}

The advent of tiny artificial intelligence (AI) accelerators, such as the Analog MAX78000~\cite{max78000}, MAX78002~\cite{max78002}, Google Coral Micro~\cite{coralMicro} and GreenWaves GAP-9~\cite{gap89} has brought AI closer to us than ever before, offering reduced latency, low power cost, and improved privacy. These accelerators, designed for microcontrollers (MCUs) with small form factors (e.g., MAX78000: 8mm$\times$8mm in Figure~\ref{fig:motivation}), are becoming integrated into wearable devices recently~\cite{balaji2023ai,omnibuds,shift,gap9hearables}, e.g., smart earbuds, patches, watches, glasses, wristband, and shoes. This integration transforms wearable devices from smartphone-dependent accessories---used merely for data collection or notification alerts---into AI-capable devices.

With the proliferation of wearable devices, it is natural to expect that personal computing environments on the body will form a network of these AI-capable wearables~\cite{min2024ai}. This enables a new class of \emph{on-body AI apps} to emerge, which enhance app functionality by leveraging on-device AI capabilities with diverse sensing capabilities and interaction methods available across wearables. These apps, running on wearable devices, continuously monitor a variety of user contexts and proactively provide context-aware services to users directly via diverse wearable interfaces. Figure~\ref{fig:motivation} shows an example scenario with three concurrent on-body AI apps: (i) \textit{memory augmentation} detects greeting words using a smartwatch microphone and captures nearby faces using glasses-mounted cameras, (ii) \textit{attention alert} monitors surrounding visual events through glasses and provides haptic alerts on a ring, and (iii) \textit{personal fitness coach} analyzes exercise routines and vital signs on a smartwatch and smart shoes, providing auditory feedback via earbuds.

% \rev{This computing environment, surrounded by AI-capable ultra-low-power wearable devices, necessitates the emergence of a novel multi-sensory AI-native system to enable the collaboration of tiny AI accelerators. While there have been many attempts in the research and industry domain to use ultra-low-power devices such as MCUs for wearable service, providing system supports to these devices have received less attention and their usage is still limited to supporting pre-define, hardcoded capabilities due to lack of system support. Such firmware-based static operation also makes it challenging to address the dynamics of wearable computing and to benefit from the collaboration of distributed devices. By envisioning the development of a multi-sensory, AI-native system, we believe it can support multiple on-body AI apps with enhanced performance efficiently by performing system-driven active collaboration of distributed AI accelerators. The system can also dynamically select the most suitable devices to execute tasks for sensing, model execution, and interaction. By strategically distributing model workloads across different AI accelerators, the system can prevent resource conflicts and ensure smooth operation, improving both individual app performance and overall system efficiency. Additionally, by splitting concurrent models and running the split chunks over multiple AI accelerators, the system maximizes processing capacity, supporting more models simultaneously and accommodating larger models that a single AI accelerator cannot handle.}

This computing environment, surrounded by AI-capable wearable devices, presents exciting opportunities for the runtime system. By leveraging the runtime collaboration of AI accelerators, it can support on-body AI apps with enhanced performance efficiently. The system can dynamically select the most suitable devices to execute tasks for sensing, model execution, and interaction. By strategically distributing model workloads across different AI accelerators, the system can prevent resource conflicts and ensure smooth operation, improving both individual app performance and overall system efficiency. Additionally, by splitting concurrent models and running the split chunks over multiple AI accelerators, the system maximizes processing capacity, supporting more models simultaneously and accommodating larger models that a single AI accelerator cannot handle.

However, it is not straightforward to realize these benefits. In the current paradigm for multi-wearable programming, task-device assignment decisions are often made at the development time. These individual app-level decisions may not be optimal at runtime due to the dynamic nature of wearable devices and dependencies on resource usage between concurrent apps. One may consider existing model partitioning techniques~\cite{kang2017neurosurgeon, dads2019dynamic, qdmp2020towards, laskaridis2020spinn, jeong2018ionn, cas2021context, banitalebi2021autosplit}. While they support dynamic model splitting, they mainly focus on optimizing the splitting decision for \textit{a single AI model}, thereby lacking a holistic view that considers (i) interdependency among different tasks---sensing, model inference, interaction on distributed devices---within an app and (ii) resource conflict across concurrent apps. Additionally, no existing partitioning work has been built for tiny AI accelerators yet, making it difficult to directly adopt them for on-body AI apps. \rev{Note that such dynamic challenges can be addressed by sending raw sensor and offloading all processing tasks to smartphones, which can be assumed to be always available. However, this incurs high communication costs to both wearable devices and the smartphone, thereby leading to increased power consumption and degraded inference throughput of AI models, as discussed in \S\ref{subsec:collaboration_benefit}.}

We present \system{}, a first-of-its-kind runtime system designed to efficiently support on-body AI apps on tiny AI accelerator-equipped wearable devices. At its core, \system{} provides runtime orchestration that makes holistic decisions of task-device assignments for concurrent apps over distributed AI accelerators. To this end, \system{} decouples task-device assignment decisions from app logic by providing device-agnostic programming interfaces. Then, \system{} dynamically creates various execution plans (mapping logical tasks to physical devices) for each app, considering available resources, communication over devices, model splitting options, etc. \system{} selects the optimal set of execution plans to maximize the system-wide performance. In this paper, we focus on maximizing the execution throughput of AI apps, but other objectives can be adopted as well. 

To this end, \system{} has four key components. (i) \system{} generates holistic collaboration plans, each representing the mapping of all logical tasks in concurrent apps to physical resources, by holistically considering resource demands from concurrent apps and resource availability of wearables. (ii) For efficient decision-making on resource-constrained MCUs, we propose a data intensity-aware pipeline accumulation method that helps \system{} reduce the exponential search cost to linear cost while providing comparable performance to the complete search. (iii) We devise a novel latency estimation model designed for tiny AI accelerators which is used to estimate and compare the throughput of holistic collaboration plans. (vi) After deployment, \system{} further improves model throughput via an adaptive task parallelization scheduler. It reduces the latency of a selected collaboration plan by maximizing parallelization opportunities of concurrent apps at runtime across distributed computation units on wearables (e.g., processors, AI accelerators, wireless chips).

We prototyped \system{} on two tiny AI accelerator platforms, MAX78000 and MAX78002. We compare \system{} with 7 baselines including state-of-the-art model partitioning techniques~\cite{kang2017neurosurgeon, dads2019dynamic, qdmp2020towards, laskaridis2020spinn, jeong2018ionn, cas2021context, banitalebi2021autosplit}. Our extensive evaluation with 8 AI models shows that \system{} consistently outperforms the baselines, with on average 23.0$\times$ throughput gain across various scenarios, while reducing latency by 73.9\% and power consumption by 15.8\%. Our in-depth experiments also show that \system{} effectively adapts to various runtime environment changes: the number of devices, the number of pipelines, heterogeneous device resources, different source and target mapping, and different objectives.

\section{Background \& Motivation}

\subsection{Tiny AI Accelerators}

The integration of AI accelerators into MCUs represents a significant move towards distributed, on-device AI, ensuring enhanced user privacy and minimal latency. Although a number of tiny-scale accelerators have been proposed recently, very few products are commercially available and provide access and control over their underlying operations. In this paper, we chose Analog MAX78000~\cite{max78000} and MAX78002~\cite{max78002} as our primary platforms since they are the most widely used tiny AI accelerator research platforms~\cite{moosmann2023tinyissimoyolo, ruegg2023kp2dtiny, bakar2022protean, caronti2023fine, moss2022ultra, gong2024dex, jeon2025tinymem, jang2024thermal, huang2024energy} owing to the disclosed hardware details and open-source tools, enabling in-depth analysis and modification of operations.

Both MAX78000 and MAX78002 have an Arm Cortex-M4 processor with different memory capacities; 512 KB of flash and 128 KB of SRAM on MAX78000 and 2.5 MB and 384 KB on MAX78002. For acceleration, they feature a convolutional neural network (CNN) accelerator that contains 64 CNN processors specially designed for parallel convolutional operations at ultra-low power. The CNN accelerator has the dedicated memory; 512 KB of data memory, 442 KB of weight memory, and 2 KB of bias memory on MAX78000 and 1.3 MB, 2 MB, and 8 KB on MAX78002, respectively. 

\begin{figure}[t!]
\vspace{-0.1in}
    \centering
    \includegraphics[width=0.99\linewidth]{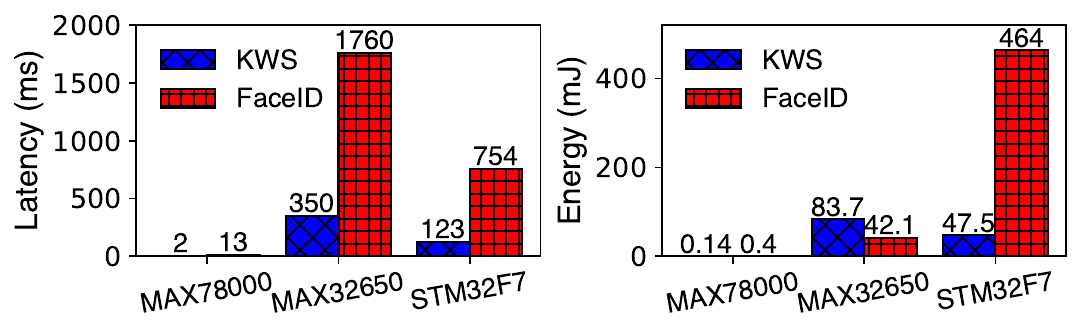}
    \vspace{-0.5cm}
    \caption{Performance comparison between AI accelerator (MAX78000) and MCUs (MAX32650 and STM32F7).}
    \vspace{-0.35cm}
    \label{fig:mcu_perf}
\end{figure}

Recent benchmark study~\cite{max78000benchmark,moss2022ultra} quantifies the MAX78000's superior performance in terms of latency and energy cost. Figure~\ref{fig:mcu_perf} shows MAX78000 significantly outperforms a conventional MCU, MAX32650 with Cortex-M4 at 120~MHz~\cite{max32650} and even a high-performance MCU, STM32F7 with Cortex-M7 at 216 MHz~\cite{stm32f7} in key AI models. Latency for keyword spotting (KWS) is reduced to 2.0 ms compared to 350 ms and 123 ms for MAX32650 and STM32F7, respectively. Energy efficiency is similarly enhanced, with MAX78000 consuming merely 0.40 mJ for face detection (FaceID), in contrast to 42.1 mJ and 464 mJ consumed by MAX32650 and STM32F7.

\subsection{AI Accelerator Collaboration}~\label{subsec:collaboration_benefit}
With wearable devices integrating tiny AI accelerators, we foresee a proliferation in on-body AI apps running across distributed wearables (Figure~\ref{fig:motivation}). These environments offer opportunities to enhance performance through runtime collaboration of AI accelerators as follows.

\textbf{Dynamic device selection:} The system can dynamically allocate tasks such as sensing, model execution, and interaction to the most suitable devices based on resource availability and conditions. For example, a higher-performance AI accelerator can be chosen for an inference task as the set of available wearable devices changes.

\textbf{Strategic workload distribution:} Concurrent apps often face resource conflicts. For example, assigning multiple models to the same AI accelerator can cause out-of-resource (OOR) errors if their combined size exceeds the capacity. By distributing workloads across AI accelerators, the system prevents conflicts and ensures efficient execution of apps.

\textbf{Optimizing AI accelerator utilization:} Tiny AI accelerators are usually optimized for single-model support. By splitting and distributing models across multiple AI accelerators, the system can maximize processing capacity. This approach supports more models simultaneously and accommodates large models that cannot fit into a single AI accelerator.

\begin{figure}[t!]
\vspace{-0.1in}
    \centering
    \includegraphics[width=0.99\linewidth]{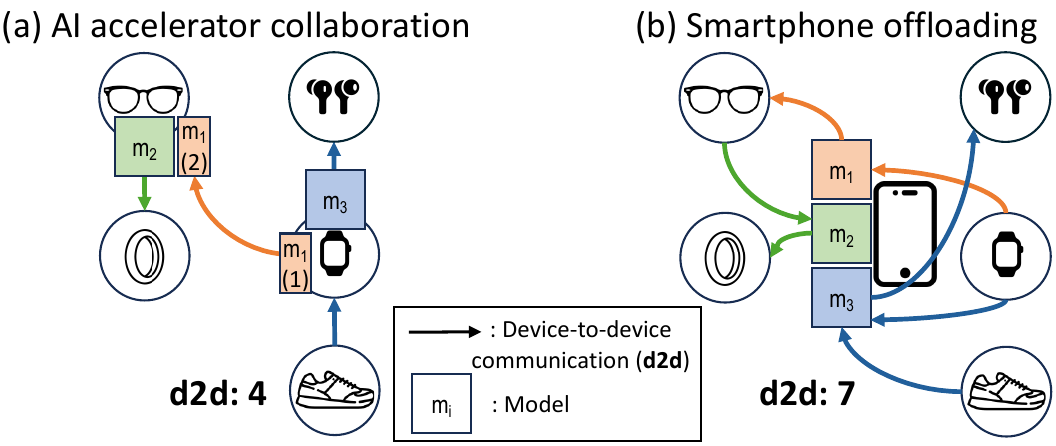}
    \vspace{-0.3cm}
    \caption{Comparison between (a) AI accelerator collaboration via \system{} and (b) phone offloading.
    }
    \vspace{-0.1cm}
    \label{fig:offloading}
\end{figure}

\begin{figure}[t!]
% \vspace{-0.1in}
    \centering
    \includegraphics[width=0.95\linewidth]{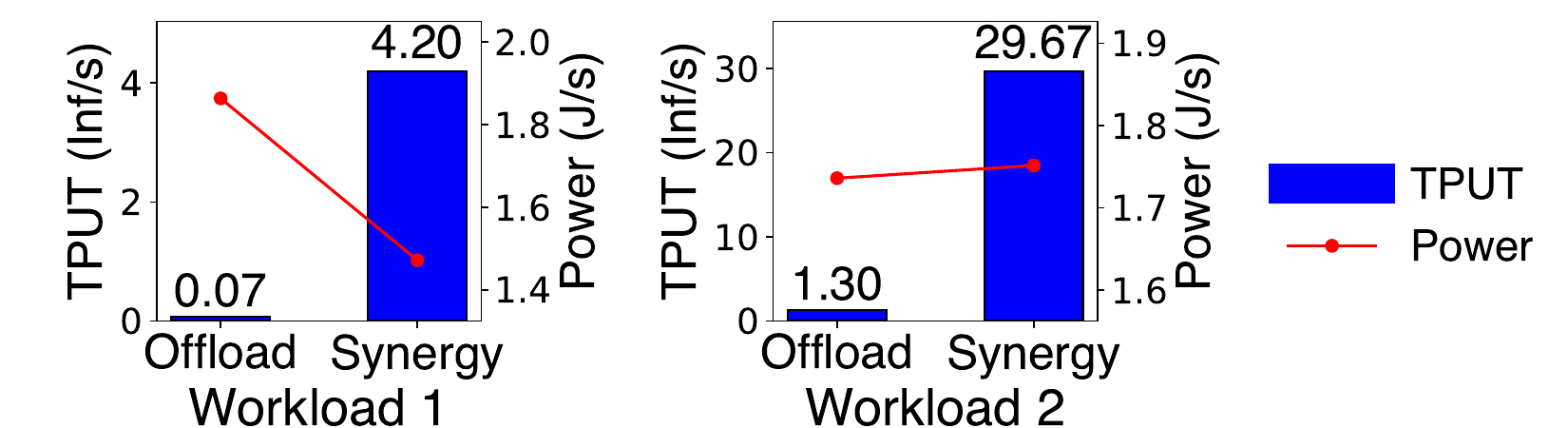}
%    \vspace{-0.3cm}
    \caption{Comparison of \system{} and phone offloading.}
    % Throughput and power comparison of \system{} and offloading. }
    \vspace{-0.3cm}
    \label{fig:offloading_vs_synergy}
\end{figure}

\subsubsection{Why not offloading?} Offloading AI model execution to a smartphone might seem advantageous due to its higher processing capability. \rev{In particular, this would be true for conventional mobile apps where model execution occurs intermittently, mostly upon a user's request, and the service with the final inference output is provided on a smartphone.} 
However, this approach can be inefficient for on-body AI apps because (i) they require continuous sensing and inference for situational services and (ii) services are directly provided through wearable devices. \rev{Figure~\ref{fig:offloading} shows an illustrative comparison when the three apps introduced in \S\ref{sec:introduction} are concurrently running. While the AI accelerator collaboration can support the concurrent execution of three apps with four device-to-device (d2d) communications as shown in Figure~\ref{fig:offloading} (a), smartphone offloading requires seven d2d communications as shown in Figure~\ref{fig:offloading} (b), as the data must pass through a smartphone regardless of the data path. Thus, such offloading incurs additional data communication and results in increased latency and energy costs. These additional transmission costs are unnecessary when apps run solely on wearables.} 
% As shown in Figure~\ref{fig:offloading}, offloading incurs additional data communication, resulting in increased latency and energy costs. 
% These costs are unnecessary when apps run solely on wearables. 

Continuous transmission of raw sensor data to a smartphone also causes communication and energy bottlenecks, affecting both smartphones and wearables.
We compared the total throughput of model execution using Workload 1 and 2 from Section~\ref{sec:evaluation:overall}. By eliminating unnecessary links, \system{} improves throughput by 57.7$\times$ and 28.8$\times$ compared to smartphone offloading, as shown in Figure~\ref{fig:offloading_vs_synergy}. Despite frequent model execution, \system{} consumes less or comparable power due to avoiding energy-intensive data communication.

\rev{Based on these findings, we can conclude that offloading (sending raw sensor data to a smartphone) is not effective when (i) the size of raw sensor data is larger than that of the intermediate or final results, (ii) the final results need to be delivered to wearable devices for the service, and (iii) the model needs to be executed frequently for situational and proactive services, which is the case for on-body AI apps.}

\section{\system{} Design}

% \subsection{Holistic Orchestration for AI Accelerator Collaboration}

% \subsubsection{Limitations of Existing Methods}

% In the current paradigm for multi-wearable programming, task-device assignments are predetermined in the app logic at development time. This \textit{app-level assignment decision} has limitations in environments with concurrent AI apps on heterogeneous wearables with varying availability. For example, a memory augmentation app using a smartwatch's AI accelerator would underperform or fail if other apps compete for the same accelerator or if the smartwatch’s battery depletes, even when other devices have available accelerators. Leveraging multiple AI accelerators is also challenging because developers lack visibility into available accelerators and their capabilities at development time.

\begin{figure}[t]
    \centering
    \includegraphics[width=0.95\linewidth]{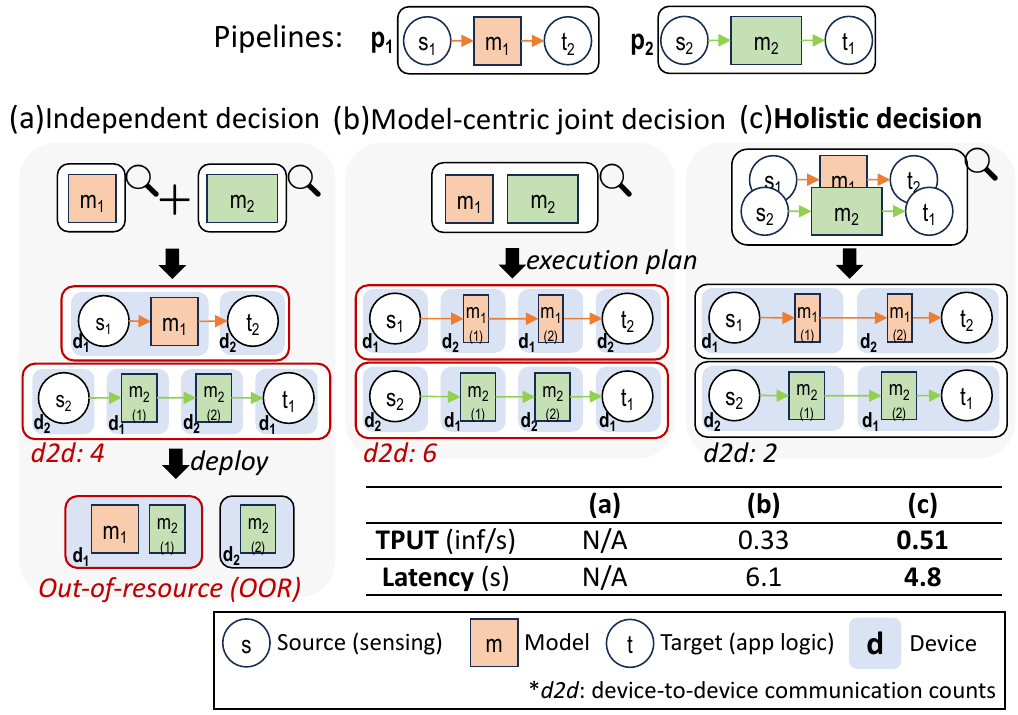}
%    \vspace{-0.1in}
    \caption{Comparison of decision approaches.
    }
    \vspace{-0.1in}
    \label{fig:method_comparison}
\end{figure}

\rev{
\subsection{Limitations of Existing Approaches}
In the current multi-wearable programming paradigm, task-device assignments are predetermined at development time, limiting adaptability in dynamic environments with concurrent AI apps and heterogeneous wearables. This app-level assignment decision poses several challenges:
\begin{itemize}
    \item \textbf{Lack of adaptability to resource contention:} If a memory augmentation app relies on a smartwatch’s AI accelerator, it may underperform or fail due to competition from other apps or battery depletion, even when other devices have available accelerators.
    \item \textbf{Limited visibility into available resources:} Developers lack real-time awareness of available AI accelerators and their capabilities, making it difficult to leverage multiple devices efficiently.
\end{itemize}
Existing model partitioning techniques~\cite{kang2017neurosurgeon, jeong2018ionn, dads2019dynamic, qdmp2020towards, laskaridis2020spinn, cas2021context, banitalebi2021autosplit} primarily focus on offloading computation to powerful devices (e.g., smartphones or cloud). While they could be adapted for distributed on-body AI accelerators, they have fundamental limitations:

\begin{itemize}
    \item \textbf{Single-model optimization:} These techniques optimize splitting decisions for individual models, without considering: (i) Resource dependencies between concurrent apps. (ii) Interdependencies between model inference and other tasks (e.g., sensing and interaction). 
    \item \textbf{Incompatibility with tiny AI accelerators:} No existing method is specifically designed to partition models onto tiny AI accelerators, leading to inaccurate performance modeling in resource-constrained environments.
    \item \textbf{Independent decision-making:} Existing partitioning methods split and deploy each model independently, ignoring conflicts among models.  
    This can lead to resource contention and out-of-resource (OOR) failures, particularly for memory-constrained accelerators.  
    \textit{Example:} As shown in Figure~\ref{fig:method_comparison} (a), if two apps deploy model chunks to the same device ($d_1$), the full model of $p_1$'s $m_1$ and the partial model of $p_2$'s $m_2$ may exceed $d_1$’s memory capacity, leading to OOR failures.
    \item \textbf{Model-centric joint decision:} Some methods extend model partitioning to account for memory constraints, filtering out unsupported cases.  
    However, this model-centric view still remains suboptimal, as it ignores sensing and interaction costs in an app’s end-to-end pipeline.  
    \textit{Example:} As illustrated in Figure~\ref{fig:method_comparison} (b), even when split models of $m_1$ and $m_2$ avoid memory conflicts, long-distance execution between source and target devices increases communication overhead, degrading overall performance.
\end{itemize}

These limitations highlight the need for a holistic, system-driven approach that considers concurrent workloads, device heterogeneity, and end-to-end execution costs.
}

\rev{

\subsection{Our Approach: AI Accelerator Collaboration with a Holistic View}

We propose shifting from app-level independent decision-making to system-driven collaboration in environments with concurrent AI applications on distributed AI accelerator-equipped wearables. Our approach introduces the following key innovations:
\begin{itemize}
    \item \textbf{Decoupled task-device assignment}: Existing approaches tightly couple task execution with app logic, limiting flexibility. We decouple task-device assignment, allowing the runtime system to gain system-wide visibility and fine-grained control over AI accelerator collaboration.
    \item \textbf{Device-agnostic programming interface}: \system{} provides a programming model (\S\ref{subsec:programming}) that enables developers to define end-to-end AI pipelines without specifying target devices. This abstraction facilitates dynamic task placement, adapting execution based on real-time resource availability.
    \item \textbf{Holistic collaboration and runtime optimization}: Unlike existing methods that optimize resource allocation per app, \system{} takes a holistic approach, considering both intra-app and inter-app dependencies. It constructs a system-wide collaboration plan by leveraging direct knowledge of concurrent workloads, resource availability, and task dependencies. The runtime dynamically assigns tasks to devices based on their capabilities and splits AI models across distributed AI accelerators to maximize efficiency.
    \item \textbf{End-to-end performance optimization}: Beyond model partitioning, \system{} optimizes task placement by minimizing communication overhead between source, execution, and target devices. This improves throughput and reduces overall latency. Figure~\ref{fig:method_comparison} (c) illustrates an example of this holistic collaboration.
\end{itemize}
By adopting a system-driven approach, \system{} overcomes the limitations of independent per-app optimization, improving efficiency, reducing latency, and enhancing AI accelerator utilization in multi-app distributed AI environments.}

\subsection{System Scope}

\textbf{Moderator-initiated orchestration:} In our current implementation, orchestration tasks—discovering and managing devices, creating and selecting holistic collaboration plans, and deploying them to devices—are managed by an external moderator, such as a smartphone. This is due to the limited computing capabilities of MCUs and their lower accessibility compared to smartphones, as users may not always wear their wearables. Orchestration is needed only when there is a change in apps or device availability. Once set up, runtime model inference operates independently on wearable devices. We envision a shift towards a more decentralized approach, embedding these orchestration capabilities into powerful wearables, facilitating self-sufficiency in wearable AI systems and reducing the need for external devices.

\textbf{Target metric:} To execute multiple models across distributed devices, various system-wide objectives can be considered, such as maximizing inference throughput, minimizing latency, and reducing energy consumption. This paper focuses on maximizing overall system-wide model inference throughput, a key quality of service metric in AI apps. To ensure fairness among models, we merge multiple app pipelines into a unified one and maximize its execution per second. Simply maximizing total throughput—the number of model executions per second—could let lower-latency models monopolize resources, causing fairness issues. This target metric can be replaced with other objectives, such as minimizing latency or energy costs. Results are reported in \S\ref{sec:evaluation:target_metric}.

\rev{
\textbf{Potential synergy with smartphones:} While \system{} primarily focuses on tiny AI accelerators on wearable devices and their collaboration, and we show its benefits over phone offloading for on-body AI apps in \S\ref{subsec:collaboration_benefit}, incorporating smartphones could enhance system performance and flexibility in other scenarios. In particular, a hybrid approach leveraging both wearables and smartphones can be advantageous for cases where large AI models need to run with body signals, e.g., foundation models with IMU data~\cite{das2024primus} and optical physiological signals~\cite{pillai2025papagei}. Another example would be when the final inference outcome is served using a mobile UI on a smartphone, thereby requiring no transmission overhead from a smartphone to wearable devices. By integrating the synergy with smartphones, \system{} could extend its applicability to a broader range of AI-driven wearable apps while maintaining seamless user experiences and efficient on-body computing. We leave this as future work.
}
\section{\rev{\system{} Runtime Technical Components}}\label{sec:runtime} 

\subsection{Overview}
\label{sec:overview}
\rev{
This section provides a high-level overview of our approach. We first summarize the key technical novelties that distinguish \system{} from existing solutions, followed by a step-by-step description of \system{}’s operational flow in multi-wearable environments.

\subsubsection{Summary of Technical Novelties}
\label{subsec:technical_novelties}

\system{} introduces several innovations that improve performance and flexibility for on-body AI applications:

\begin{itemize}
    \item \textbf{Device-agnostic programming interface:}  
    An abstracted interface for building on-body AI pipelines without specifying target devices at design time.

    \item \textbf{Holistic collaboration planning:}  
    A global approach that considers all concurrent apps together, supporting multi-model splitting and effective resource sharing.

    \item \textbf{Progressive search space reduction:}  
    A method to mitigate the combinatorial explosion of multi-pipeline orchestration, reducing complexity from exponential to near-linear in the number of pipelines.

    \item \textbf{Clock-cycle-based latency modeling:}  
    A precise inference-latency estimator that captures per-layer execution time on tiny AI accelerators by analyzing hardware clock cycles.

    \item \textbf{Online throughput estimation:}  
    A lightweight scheme for dynamically evaluating end-to-end performance in real time, enabling adaptive selection of the best collaboration plan.

    \item \textbf{Adaptive task parallelization:}  
    Mechanisms that exploit idle computation units by overlapping tasks across multiple pipelines (inter-pipeline) and consecutive runs (inter-run).
\end{itemize}

These novelties overcome limitations of per-app approaches by jointly managing on-body resources. Next, we describe how \system{} integrates these techniques into its operational flow.

\subsubsection{Operational Flow}
\label{subsec:operational_flow}

Figure~\ref{fig:operation_flow} depicts the runtime stages of \system{}:

\begin{figure}[t!]
    \centering
    \vspace{-0.05in}
    \includegraphics[width=1.0\linewidth]{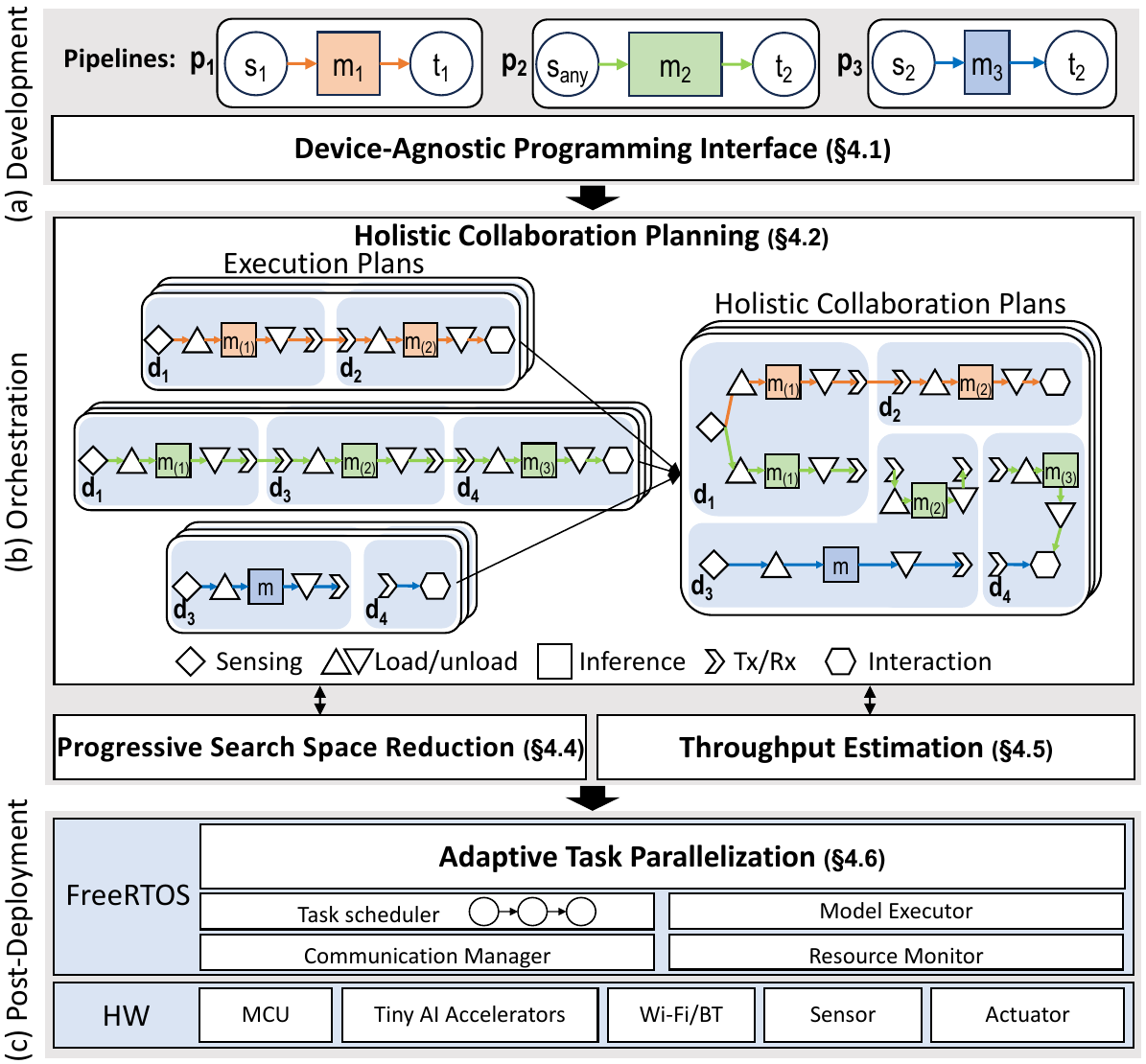}
    \vspace{-0.4cm}
    \caption{An illustration of \system{}'s operational flow.}
    \vspace{-0.5cm}
    \label{fig:operation_flow}
\end{figure}

 \textbf{App development \& installation stage}:
On-body AI apps are developed using the \textit{device-agnostic programming interface} (\S\ref{subsec:programming}). Developers only need to specify the pipeline’s logical tasks without worrying about which wearables will execute them. Once the app is installed, the pipeline description is registered with a moderator that manages device assignments.

 \textbf{Orchestration stage}:
When running apps or available devices change, \system{} triggers \textit{holistic orchestration} by generating holistic collaboration plans (\S\ref{sec:runtime:holistic}). It maps all tasks in concurrent apps to devices by matching task requirements with device capabilities and potentially splitting model tasks across distributed AI accelerators. Because exhaustively searching for the best split and device assignment can be computationally intractable on resource-constrained MCUs, \system{} applies \textit{progressive search space reduction} (\S\ref{sec:runtime:planning}) to prune infeasible or suboptimal plans. Once feasible candidates are formed, \system{} employs a \textit{novel latency estimation model} and an \textit{online throughput estimation} method (\S\ref{sec:runtime:policy}) to select the plan expected to yield the highest overall throughput.

 \textbf{Post-deployment stage}:
As multiple pipelines execute under a single holistic collaboration plan, the entire execution cycle completes when all pipelines finish. This can create idle periods on certain computation units (e.g., MCUs, AI accelerators, wireless chips). To further boost throughput at runtime, \system{} incorporates \textit{adaptive task parallelization} (\S\ref{sec:runtime:deployment}), which overlaps tasks across pipelines (inter-pipeline) and across multiple runs (inter-run) to reduce idle time and improve overall system performance.

By coordinating these stages under a single, unified framework, \system{} effectively balances task distribution, model-splitting decisions, and parallelization strategies to achieve high performance and resource utilization across multiple on-body AI accelerators.
}

\subsection{Device-Agnostic Programming Interface}~\label{subsec:programming}
We structure on-body AI apps as a directed acyclic graph (DAG) of tasks for several reasons. First, DAGs align well with the architecture of many AI-centered apps. Second, this structure represents data and device dependencies among tasks, providing \system{} visibility and controllability of concurrent apps. Third, DAGs inherently abstract app logic as a series of tasks, simplifying the division and allocation of these tasks across distributed devices.

In this work, we categorize tasks into three types: \textit{sensing}, \textit{model}, and \textit{interaction}. Sensing tasks are specified with requirements such as sensor type, resolution, and position. Interaction tasks are described with requirements like interface type and physical location. Currently, \system{} supports two types of requirements: designated device and sensor type for sensing tasks, and designated device or interface type for interaction tasks. For model tasks, apps specify an AI model to be executed. For example, a memory augmentation app pipeline could be \textit{(microphone, KeywordSpotting, camera on glasses)}, and an attention alert app could be \textit{(camera on glasses, MobileNet, haptic)}. In this paper, we support three tasks for a pipeline, but it can be expanded to a DAG, e.g., conditional inference or fusion models.

\subsection{Holistic Collaboration Planning~\label{sec:runtime:holistic}}

\textbf{Execution plan creation:} As an initial step to exploit available collaboration options, \system{} creates various \emph{execution plans} for each app to run over distributed devices. An execution plan abstracts the task-device assignment of a pipeline, serving as the basis for holistic orchestration. The key to obtaining diverse execution plans from each pipeline is (i) exploring various combinations of model splitting layers across distributed AI accelerators and (ii) flexibly mapping source and target tasks to suitable devices.

An execution plan is defined as a sequence of tuples, each containing a task and an assigned device (Figure~\ref{fig:operation_flow} (b)). \system{} supports seven task types: (i) sensing, (ii) data loading to accelerator memory, (iii) (partial) model inference, (vi) data unloading from accelerator memory, \rev{(v) Tx (transmitting data), (vi) Rx (receiving data)}, and (vii) interaction. For two wearables, smart glasses and a smart ring, a pipeline can be described as (\textit{camera on glasses}, \textit{EfficientNet}, \textit{haptic feedback}); EfficientNet has 29 layers. Then, one execution plan example would be [\textit{camera} $\rightarrow$ \textit{data loading} $\rightarrow$ $\textit{EfficientNet}^{0:19}$ $\rightarrow$ \textit{data unloading} $\rightarrow$ \textit{Tx to ring}] for the glasses and [\textit{Rx from glasses} $\rightarrow$ \textit{data loading} $\rightarrow$ $\textit{EfficientNet}^{19:29}$ $\rightarrow$ \textit{data unloading} $\rightarrow$ \textit{haptic}] for the smart ring, where $\textit{EfficientNet}^{i:j}$ refers to the model's layer $i$ to $j$.

\textbf{Runnable holistic collaboration plan generation:} The next step is to generate \emph{holistic collaboration plans}, each integrating execution plans from all pipelines (one execution plan per each app). It gives \system{} visibility over resource competition and dependencies across devices (see Figure~\ref{fig:operation_flow} (b)). A key consideration is filtering out plans that cause OOR cases. We consider three memory constraints: (i) weight memory, (ii) bias memory, and (iii) the maximum number of layers supported (e.g., 442 KB, 2 KB, 32 for MAX78000~\cite{max78000} and 1.3 MB, 2 MB, 128 for MAX78002~\cite{max78002}). A collaboration plan is \emph{runnable} if the total weight memory, bias memory, and number of layers for assigned tasks do not exceed the AI accelerators' capacities.

\subsection{Progressive Search Space Reduction}\label{sec:runtime:planning}

\begin{figure}[t]
    \centering
    \includegraphics[width=0.9\linewidth]{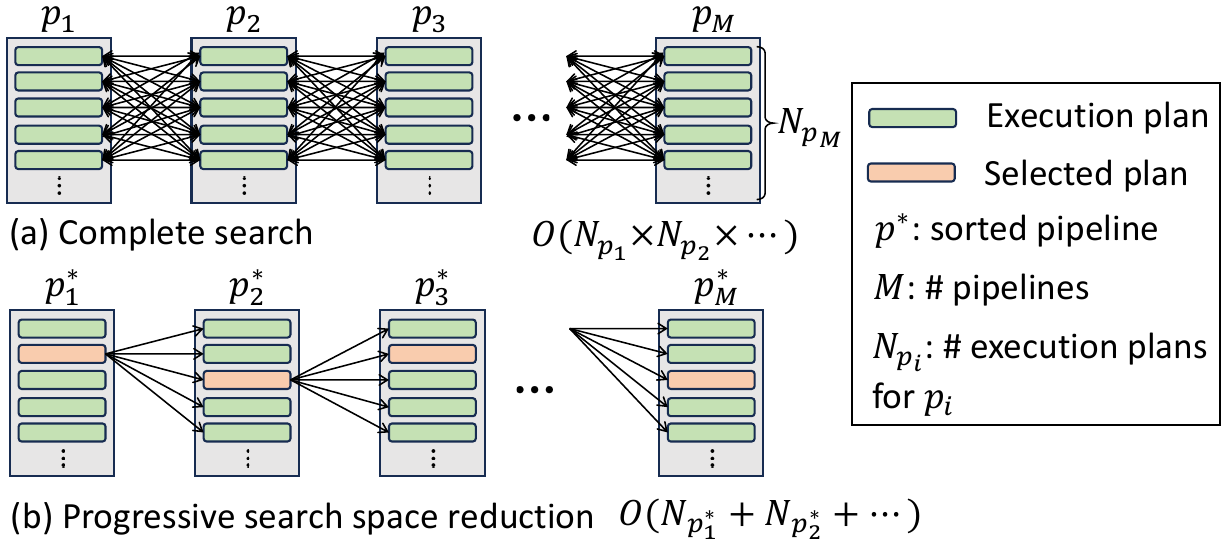}
%    \vspace{-0.4cm}
    \caption{Comparison between complete search (a) and our progressive pipeline selection (b).}
    \vspace{-0.4cm}
    \label{fig:progressive_pipeline_selection}
\end{figure}

We propose a \emph{progressive search space reduction} method to reduce the intractable search space for the holistic collaboration plan selection. For each pipeline $p$, the number of execution plans available $N_p$ can be modeled as: $N_p = \sum_{d=1}^{D} {}_D \mathrm{ P }_d \cdot {}_{L-1} \mathrm{ C }_{d-1} \cdot D^2 $, where $D$ and $L$ denote the number of available devices and model layers, respectively. \rev{${}_D \mathrm{ P }_d$ represents $d$-permutation of $D$, which means possible device orders to use, ${}_{L-1} \mathrm{ C }_{d-1}$ is $d-1$ combinations of $L-1$, which is model splitting candidates}, and $D^2$ is all source/target device mappings, respectively. This search space grows exponentially for holistic collaboration planning with concurrent pipelines, i.e., $O(N_{p_{1}} \times N_{p_{2}} \times ...)$. For instance, for three small models (a 9-layer KWS, a 14-layer SimpleNet, and a 19-layer UNet in~\S\ref{sec:exp:setting}) with three MAX78000 devices, the number of holistic collaboration plans that can be generated is $1,971 \times 4,941 \times 9,261 = 90,190,202,571$. It is evident that a complete search algorithm considering all combinations requires substantial computational overhead. To mitigate this, we propose \emph{data intensity-aware execution plan accumulation}, shown in Figure~\ref{fig:progressive_pipeline_selection}. This approach arranges execution plans in sequence via our data intensity metric and selects an execution plan for one app at a time, building on the collaboration plan integrated with previously selected execution plans from earlier apps, thereby reducing complexity to $O(N_{p_{1}}+N_{p_{2}}+...)$.
% Compared to the complete search that considers all combinations back and forth, our progressive pipeline selection reduces complexity from $O(N_{p_{1}} \times N_{p_{2}} \times ...)$ to $O(N_{p_{1}}+N_{p_{2}}+...)$. 

% \begin{algorithm}
% \caption{\system{} Multi-Pipeline Execution Planning}\label{alg:planning}
% \begin{algorithmic}[1]
% \Require{List of pipelines $P$, List of devices $D$, Metric $m$}
% \Ensure{Best holistic execution plan $E^*$}
% \State{$P \gets$ {\tt SortByDataIntensity}$(P)$, $E^* \gets \emptyset$} \Comment{\S\ref{sec:runtime:planning}}
% \For{each pipeline $p \in P$}
%     \State{$E_{p}\gets$ {\tt holisticExecutionPlans}$(p, D, {E^*})$} \Comment{\S\ref{sec:runtime:holistic}}
%     \If {$m==throughput$} \Comment{\S\ref{sec:runtime:policy}}
%     \State{$E^* \gets \argmax_{e_i \in E_{p}}$ {\tt Throughput}$(E^*, e_i)$} 
%     \ElsIf {$m==latency$}
%     \State{$E^* \gets \argmin_{e_i \in E_{p}}$ {\tt Latency}$(E^*, e_i)$}
%     \EndIf
% \EndFor
% % \State{\Return $E^*$}
% \end{algorithmic}
% \end{algorithm}

\begin{figure}[t]
    \centering
    \includegraphics[width=1.0\linewidth]{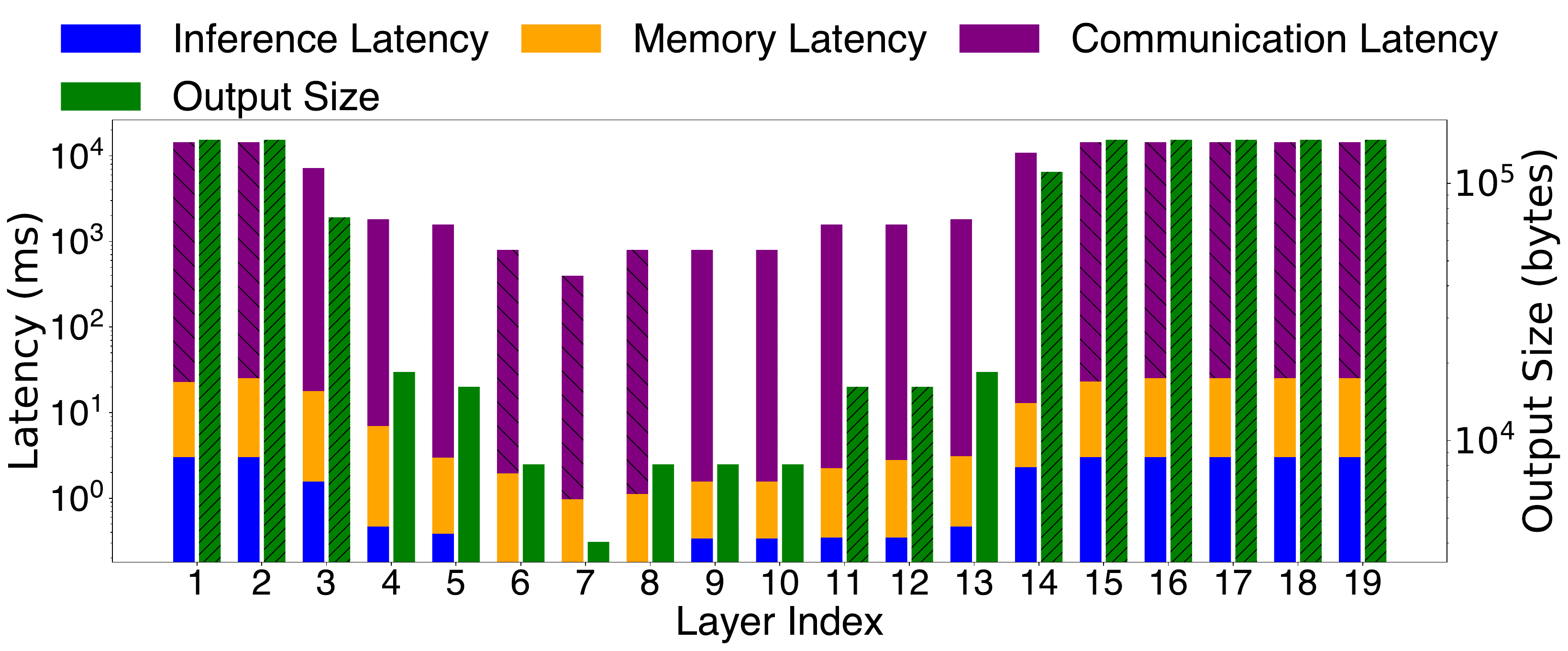}
%    \vspace{-0.8cm}
    \caption{Layer-wise latency analysis for UNet. The y-axis is in the logarithmic scale. \rev{Since the scale is logarithmic, the total height of each stacked bar represents the geometric sum of its components, making small values appear relatively smaller while highlighting differences across orders of magnitude.}}
    \vspace{-0.4cm}
    \label{fig:layer_latency}
\end{figure}

\textbf{Pipeline prioritization:} The key to data intensity-aware execution plan accumulation is determining the sequence of pipelines. \system{} prioritizes data-intensive pipelines for execution plan selection, meaning pipelines that require the transmission of larger sensing data and intermediate outputs. The intuition is that optimizing data-intensive pipelines yields greater overall benefits for the system. To understand the relationship between data intensity (i.e., data size) and latency, we conducted a preliminary measurement. Figure~\ref{fig:layer_latency} shows layer-wise latency for inference, memory, and communication, alongside output data size using UNet~\cite{ronneberger2015u} as an example. Compared to inference latency (1.5 ms), both memory latency (10.6 ms) and communication latency (6869.1 ms) are significantly higher, being $7\times$ and $4579\times$ greater, respectively. This latency gap between inference and communication is notable in tiny AI accelerators designed for fast inference. Moreover, the output size varies significantly across different layers, with the lowest layer latency at 4426.2 ms and the highest at 161864.5 ms, indicating a $36\times$ difference. These findings suggest that (i) the end-to-end latency of a pipeline varies greatly depending on the model split in the execution plan, (ii) latency is strongly related to the size of data exchanged, and (iii) prioritizing data-intensive pipelines could offer more optimization opportunities due to fewer resource conflicts with other pipelines.

\textbf{Pipeline arrangement and accumulation:} We define a pipeline's data intensity as the average data size of transmission over all execution plans. Specifically, given the input size ${\tt In}^{\tt size}$ and output size ${\tt Out}^{\tt size}_{l}$ for each layer $l \in L$, the data intensity of a pipeline is defined as: $({\tt In}^{\tt size} + \sum_{l}{\tt Out}^{\tt size}_{l})/(L+1)$. The pipelines are sorted in descending order with this metric. Then, for each pipeline, \system{} generates runnable holistic collaboration plans by combining each execution plan of the current pipeline with the previously selected execution plans from earlier pipelines. Then, \system{} selects the execution plan of which holistic collaboration plan is expected to have the highest system-wide throughput based on our online throughput estimation~(\S\ref{sec:runtime:policy}). 

\begin{figure}[t]
    \centering
    \includegraphics[width=0.95\linewidth]{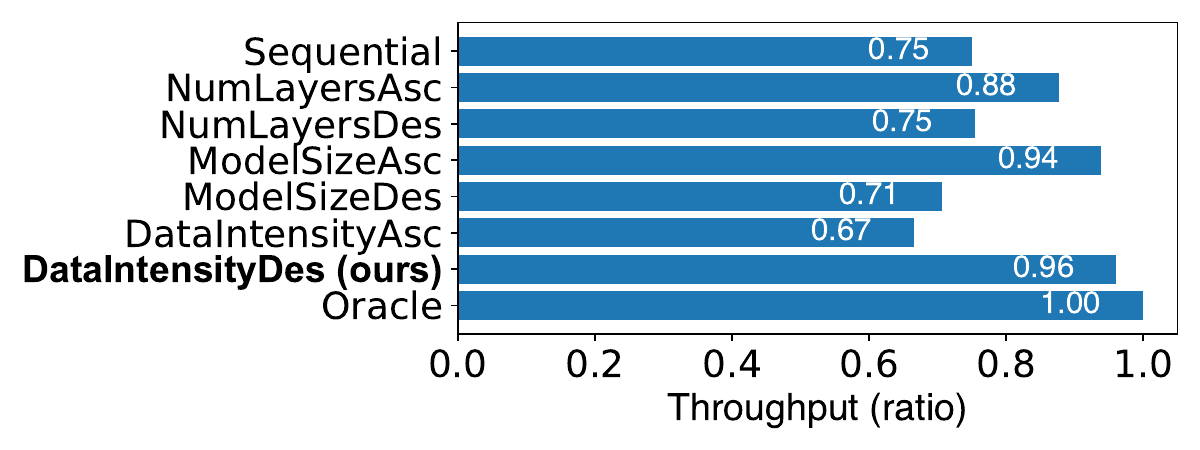}
%    \vspace{-0.4cm}
    \caption{Comparison among different pipeline prioritization strategies against complete search (Oracle).}
    \label{fig:oracle_vs_synergy}
    \vspace{-0.4cm}

\end{figure}

\textbf{Comparison with other prioritization methods:} We investigate (i) how significantly our selection reduces the search space compared to a complete search, and (ii) how close our prioritization method is to the optimal selection compared to other alternatives. We compare \system{} with seven baselines: the complete search (Oracle), ascending order of data intensity (DataIntensityAsc), descending/ascending order of model size (ModelSizeDes/ModelSizeAsc), and descending/ascending order of the number of model layers (NumLayersDes/NumLayersAsc), and without prioritization (Sequential). Except for Oracle, the six baselines use the execution plan accumulation method but with different prioritization objectives. In this experiment, we assume three concurrent pipelines with two MAX78000 devices and consider all possible combinations of three pipelines out of eight (Table~\ref{tab:models}). Figure~\ref{fig:oracle_vs_synergy} shows the relative throughput ratio compared to Oracle by averaging the throughput over all pipeline combinations. The results validate our design choice. \system{} outperforms the other alternatives, showing only a 3.9\% degradation in overall throughput compared to the complete search. \rev{Additionally, our execution plan accumulation approach results in a 5576$\times$ reduction in search space compared to Oracle. Note that the other prioritization methods have the same search space as ours because they are all based on our progressive pipeline selection.}
% Additionally, the execution plan accumulation approach (\system{} and other prioritizations) results in a 5576$\times$ reduction in search space compared to Oracle.

\subsection{Throughput Estimation for Distributed Tiny AI Accelerators}~\label{sec:runtime:policy} 
\begin{figure}[t]
    \centering
    \includegraphics[width=0.8\linewidth]{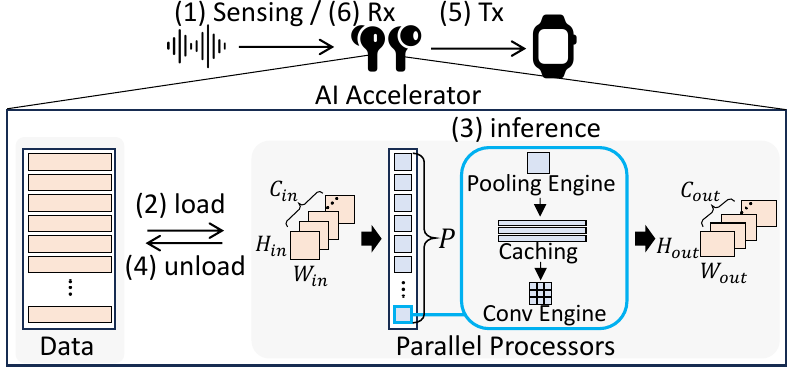}
%    \vspace{-0.4cm}
    \caption{Tasks involved in a pipeline for latency modeling ((1)-(6)) from the viewpoint of a device.}
    % We model the inference latency via clock cycles.}
    \vspace{-0.2cm}
    \label{fig:latency}
\end{figure}
For holistic collaboration plan selection, evaluating the throughput of a plan is necessary. To achieve this, we first model the execution latency for each task and estimate throughput based on this latency model. Figure~\ref{fig:latency} illustrates the operation of the five types of tasks ((1)-(6)) involved in a device. We present (i) our novel latency estimation model for model inference tasks, which reflects the unique characteristics of tiny AI accelerators (\S\ref{subsubsec:latency_ai_accelerator}), and (ii) the latency estimation for other tasks (\S\ref{subsubsec:latency_others}). We then introduce our technique for estimating the system-wide throughput of a holistic collaboration plan (\S\ref{subsubsec:throughput_estimation}).

\begin{figure}[t]
    \centering
    \includegraphics[width=0.99\linewidth]{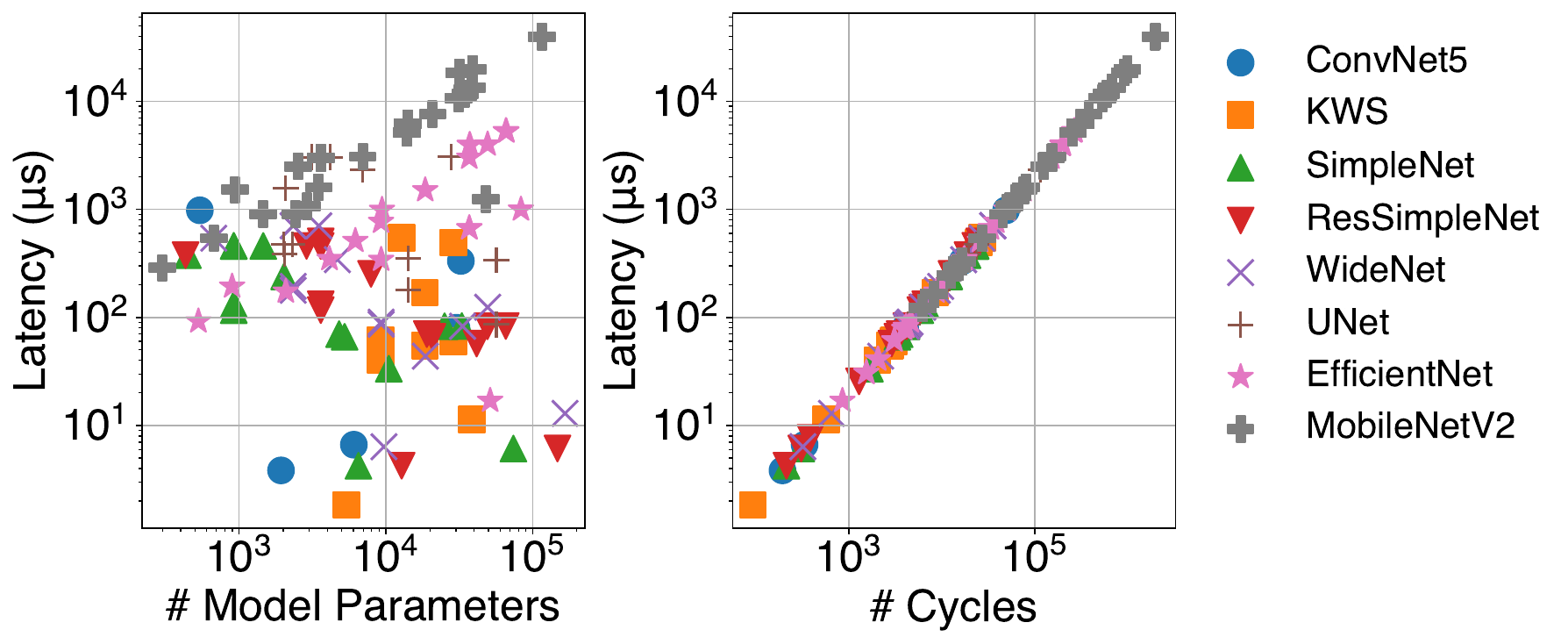}
%    \vspace{-0.4cm}
    \caption{Correlations of latency with (a) trainable parameters (left) and (b) accelerator clock cycles (right). Each dot means a different layer within a model.}
    \vspace{-0.4cm}
    \label{fig:param_lat_relation}
\end{figure}

\subsubsection{Latency Modeling for Tiny AI accelerators}~\label{subsubsec:latency_ai_accelerator} 
For latency estimation of model splitting options, existing studies rely on measurement-driven approaches: online latency measurement by running each model layer on the target device directly~\cite{laskaridis2020spinn, qdmp2020towards} or offline latency modeling with model configurations~\cite{kang2017neurosurgeon, jeong2018ionn, cas2021context}, such as learning a regression model based on the number of model parameters and latency. The first approach—direct runtime measurement for each layer—is accurate and feasible for two-tier architectures (e.g., a smartphone and a cloud server) with limited cases but entails significant overhead in our scenario involving multiple models and resource-constrained devices. The second approach—latency modeling with offline profiling—mitigates this issue but has critical limitations in our environment. First, its scalability is highly limited, requiring profiling of all possible models and AI accelerators before app deployment. Second, due to the unique characteristics of AI accelerators, existing configuration-based latency modeling may not work. Figure~\ref{fig:param_lat_relation} (a) shows the relationship between the number of model parameters and the corresponding inference latency of each layer in eight models on MAX78000. Due to hardware-level optimization in tiny AI accelerators, there is a weak correlation between these two variables.

To address these issues, we propose a \emph{clock cycle-based latency model}. Unlike measurement-driven approaches, our model calculates the number of clock cycles required for arithmetic operations of a given layer on AI accelerators, reflecting the internal operation of convolution accelerators on tiny AI accelerators. The number of clock cycles is, by design, proportional to the latency, as the model inference task runs solely on dedicated hardware (tiny AI accelerator). This approach would be inaccurate for general-purpose processors because their use is shared with other system components or apps. Specifically, the latency within the AI accelerator can be represented as follows:
\begin{equation}
    \mathcal{L}_{\tt ai\_acc} = \mathcal{L}_{\tt load} + \mathcal{L}_{\tt inf} + \mathcal{L}_{\tt unload},
\end{equation}
where $\mathcal{L}_{\tt inf}$ is the inference latency with the AI accelerator, and $\mathcal{L}_{\tt load}$/$\mathcal{L}_{\tt unload}$ is the data (un)loading latency between the processor (Arm Cortex-M4) and AI accelerator via SRAM.

\textbf{Inference latency modeling:} To model $\mathcal{L}_{\tt inf}$ ((3) in Figure~\ref{fig:param_lat_relation}), we calculate the number of clock cycles $\mathbb{C}$ at AI accelerators. We can estimate the latency of the inference by dividing the number of clock cycles per each layer $\mathbb{C}_l$ with the clock frequency $\mathcal{F}_l$ of the AI accelerator that processing layer $l$, i.e., $\mathcal{L}_{\tt inf} = \sum_{l}\frac{\mathbb{C}_{l}}{\mathcal{F}_{l}}$. Specifically, given the shape of an input $(H_{\tt in}, W_{\tt in}, C_{\tt in})$ and the corresponding output $(H_{\tt out}, W_{\tt out}, C_{\tt out})$ of a layer, representing the height, width, and number of channels respectively, the number of clock cycles for fully connected layers (MLP) and convolutional layers (CNN) under \textit{sequential} processors such as Arm Cortex-M4 are as follows: 
\begin{align}
    \hat{\mathbb{C}}_{\tt MLP} &= H_{\tt in} \cdot W_{\tt in} \cdot C_{\tt in} \cdot C_{\tt out}\\
    \hat{\mathbb{C}}_{\tt CNN} &= K^2 \cdot H_{\tt in} \cdot W_{\tt out} \cdot C_{\tt in} \cdot C_{\tt out}
\end{align}
$K$ refers to the size of the kernel. Note that tiny AI accelerators have (i) parallel computation units that parallelize operations across channels and (ii) convolution engines that process the convolution with a single clock cycle. Given the number of parallel convolutional processors, $P$, the clock cycles on AI accelerators are reduced to:
\begin{align}
    \mathbb{C}_{\tt MLP} &= H_{\tt in} \cdot W_{\tt in} \cdot \left\lceil \frac{C_{\tt in}}{P} \right\rceil \cdot C_{\tt out}\\
    \mathbb{C}_{\tt CNN} &= H_{\tt in} \cdot W_{\tt out} \cdot \left\lceil \frac{C_{\tt in}}{P} \right\rceil  \cdot C_{\tt out}
\end{align}
 % This clock-cycle-based latency modeling is especially effective for AI accelerators.

Figure~\ref{fig:param_lat_relation} (b) shows the correlation between the clock cycles of every layer in eight models and the corresponding inference latency. Unlike Figure~\ref{fig:param_lat_relation} (a), our clock cycle-based modeling shows a strong correlation by design. When we translate the number of clock cycles to the estimated latency with the clock frequency, the results show that the gap between the estimated and measured latency is less than 1\%. 

For memory operations, such as data loading and unloading between SRAM and accelerator memory ((2) and (4) in Figure~\ref{fig:param_lat_relation}), we use a measurement-driven approach. We estimate $\mathcal{L}_{\tt load}$ and $\mathcal{L}_{\tt unload}$ by profiling them with different data sizes and creating a linear regression model. Since data communication between two memories occurs through the central bus at a dedicated data transmission rate, the runtime latency of memory operations is proportional to the data size. As memory latency is not affected by the model architecture, we can model $\mathcal{L}_{\tt load}$ and $\mathcal{L}_{\tt unload}$ using a few data samples of different sizes.

\subsubsection{Latency Modeling for Other Tasks}~\label{subsubsec:latency_others} 
Besides operations on tiny AI accelerators, on-body AI apps involve sensing and communication (Rx and Tx) tasks ((1), (5), (6) in Figure~\ref{fig:param_lat_relation}). For sensing tasks, we measure the latency for camera and audio inputs with different parameters (e.g., different sampling rates) during the profiling phase and estimate the latency by matching the app's requirements to the profiles. For communication latency, similar to memory operations on tiny AI accelerators, we estimate it by dividing the data size by the bandwidth of the wireless channel. Advanced methods~\cite{kang2017neurosurgeon,qdmp2020towards} can address wireless transmission latency fluctuations, but this is beyond the scope of this paper.

\subsubsection{Throughput Estimation of Holistic Collaboration Plan}~\label{subsubsec:throughput_estimation} 
The next step is to estimate the throughput of a holistic collaboration plan (a collection of execution plans from multiple apps). To this end, \system{} initially estimates the end-to-end latency of a holistic collaboration plan based on the latency estimation of tasks in the plan. Since a holistic collaboration plan is structured as a DAG with multiple source tasks and target tasks, its end-to-end latency can be defined as the longest path from any source task to any target task; the path's length is the cumulative latency of all tasks within the path. Then, the system-wide throughput is calculated by taking the inverse of the end-to-end latency and then multiplying this value by the number of pipelines.

\begin{figure}[t]
    \centering
    \includegraphics[width=0.90\linewidth]{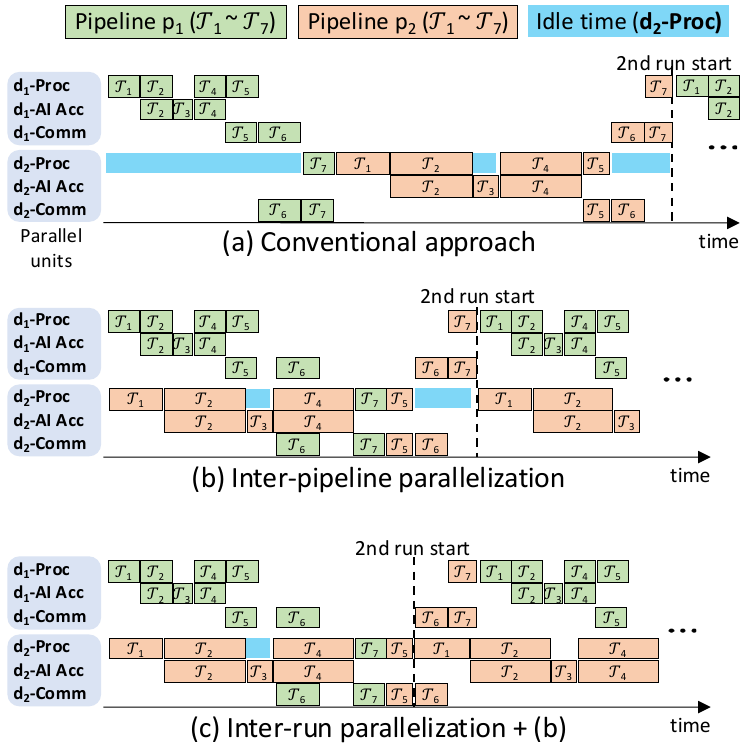}
%    \vspace{-0.4cm}
    \caption{Comparison between the conventional approach (a) and our proposed adaptive parallelization mechanisms (b, c) across two devices ($d_1$ and $d_2$).}
    \vspace{-0.3cm}
    \label{fig:parallel}
\end{figure}

\subsection{Adaptive Task Parallelization }~\label{sec:runtime:deployment}
After selecting the best holistic collaboration plan, \system{} deploys the tasks and their connection to the corresponding devices. As existing model partitioning techniques~\cite{kang2017neurosurgeon, dads2019dynamic, qdmp2020towards, laskaridis2020spinn, jeong2018ionn, cas2021context, banitalebi2021autosplit} focused on a single run for a single model, a straightforward way to execute a holistic collaboration plan with concurrent pipelines over multiple runs (continuous inference over time) is to run each pipeline sequentially---immediately run another after the completion of the preceding pipeline---in a continuous manner, as shown in Figure~\ref{fig:parallel} (a). However, since tasks in the plan runs over different computation units (processors, AI accelerator, and communication module) from distributed devices, such a sequential execution yields long idle time for unused units. The challenge is how to enhance performance by effectively utilizing these diverse computation units while keeping the sequential dependency of tasks within each pipeline.

To further enhance the system-wide throughput at runtime, we propose a method for \textit{adaptive task parallelization}, which consists of two parallelization strategies. First, \system{} maximizes parallelization opportunities by concurrently executing tasks among different pipelines (\textit{inter-pipeline parallelization}), which utilizes idle parallel computation units (Figure~\ref{fig:parallel} (b)). Note that, when multiple tasks from different pipelines are competing with the same computation units, the later-arriving tasks need to wait until the completion of the earlier tasks. Second, we extend this concept to tasks from successive runs within a pipeline (\textit{inter-run parallelization}) as shown in Figure~\ref{fig:parallel} (c); for example, the model inference in the second run does not need to wait for the completion of the pipeline in the first run when the AI accelerator is idle and the sensor data for the second run is ready.

To enable these strategies, \system{} employs separate task queues for each type of computation unit, alongside a dedicated scheduler. At runtime, on each computation unit, its scheduler is responsible for dequeuing tasks from its queue and initiating their execution. Upon task completion, the scheduler adds the next task to its corresponding queue. This structured approach maximizes the utilization of computation units, thereby significantly improving the throughput of the holistic collaboration plan at runtime.

\section{Prototype Implementation}~\label{sec:impl}
\begin{figure}[t]
    \centering
    \includegraphics[width=0.85\linewidth]{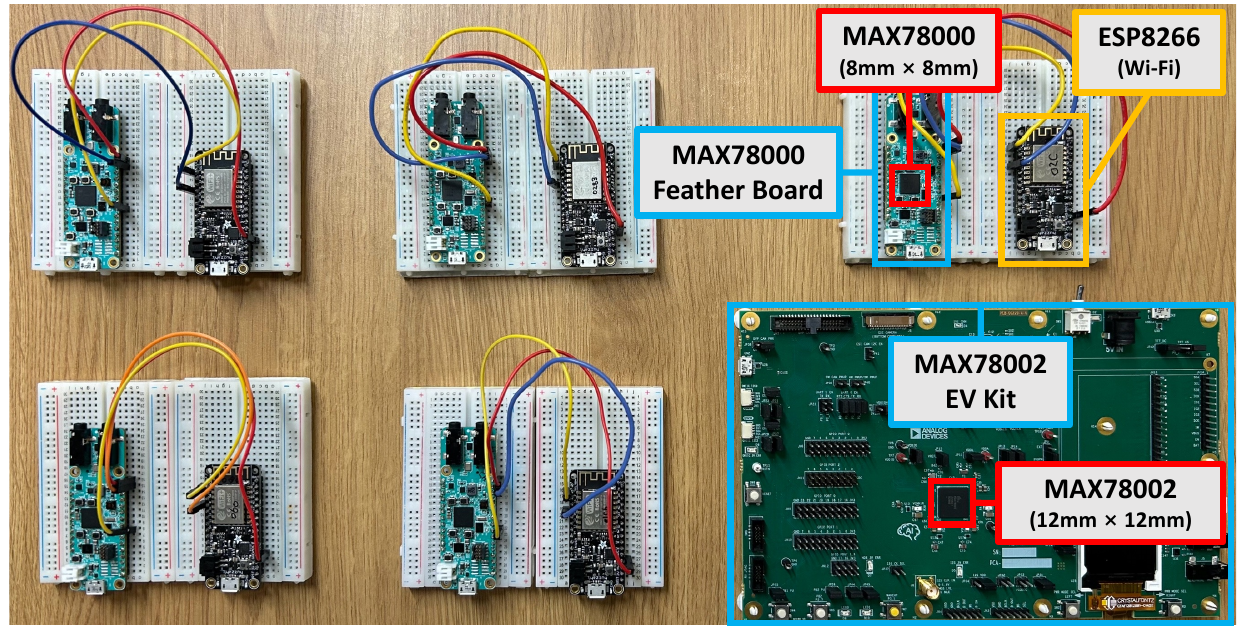}
%    \vspace{-0.15in}
    \caption{Hardware setup.}
    \vspace{-0.2in}
    \label{fig:testbed}
\end{figure}

We prototyped \system{} on the off-the-shelf MAX78000 feather board~\cite{max78000fth} and MAX78002 Evaluation Kit~\cite{max78002evkit}, which are a development platform for the MAX78000~\cite{max78000} and MAX78002~\cite{max78002}, respectively (Figure~\ref{fig:testbed}). 
Note that although the development platform is bulky, the actual size of the accelerators is tiny (e.g., 8mm$\times$8mm for MAX78000).

For wireless communication, we interfaced MAX78000/ MAX78002 with an ESP8266 Wi-Fi module~\cite{esp8266}. 
To ensure that the communication process via Wi-Fi does not disrupt the data transmissions with MAX78000/MAX78002, we have employed a round-robin scheduling algorithm on ESP8266. This mechanism is designed to alternate the data flow: transmitting data to the wireless communication channel for outward transmission and redirecting to the serial communication channel upon receiving data from the Wi-Fi interface. 
Additionally, we employ lightweight Internet Protocol a compact, standalone implementation of the TCP/IP protocol tailored for low-power devices, consuming a mere 40KB of memory.

The software system of \system{} is implemented on a FreeRTOS with C language. We abstract sensor reading, inference, and networking functionality as individual tasks, and each task is scheduled on top of the FreeRTOS scheduler. Once a joint plan is selected based on the target metric, the model is split to allow parts of the model to run on different AI accelerators. To this end, we \emph{synthesize} the partial model for a given range of layers for a designated AI accelerator. Synthesizing is the process of generating device-specific code from pre-trained models. This involves analyzing the pre-trained model and mapping inputs, weights, and outputs to memory and processor appropriately. In this study, we generate C codes from pre-trained PyTorch models for deployment on MAX78000 and MAX78002 devices.

\section{Evaluation}\label{sec:exp}

\subsection{Experimental Setup}\label{sec:exp:setting}

\subsubsection{Workload}

\begin{figure}[t]
    \centering
%    \vspace{-0.4cm}
    \includegraphics[width=0.99\linewidth]{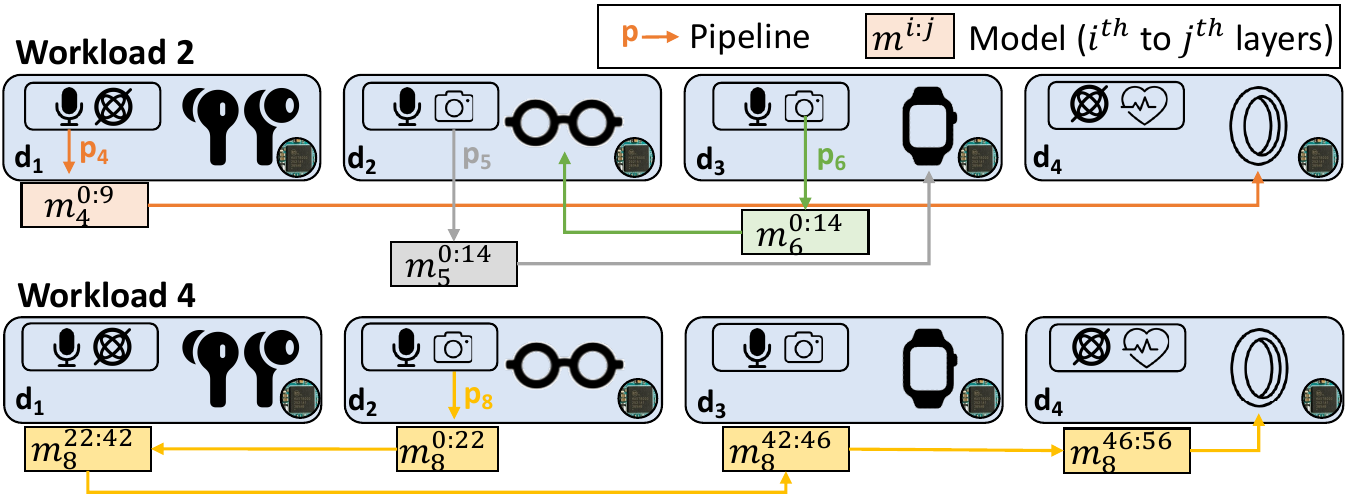}
%    \vspace{-0.4cm}
    \caption{Example execution plans (workload 2, 4).}
    \vspace{-0.4cm}
    \label{fig:workload_example}
\end{figure}

To demonstrate the effectiveness of Synergy orchestrating and executing on-body AI apps, we design four workloads with eight pipelines and eight different models (see Table~\ref{tab:models}). For the workload design, we considered two scenarios: concurrent apps, each using a different model (type, size) for different requirements (Workload 1 and 2) and a single app with a large model for enhanced capability (Workload 3 and 4). For the device setup, we used four MAX78000 devices, representing four smart wearables; an earbud, glasses, a watch, and a ring. We then assign source and target devices and a model based on the app scenarios. Figure~\ref{fig:workload_example} shows example execution plans of Workload 2 and 4. Pipeline 4 (keyword spotting), captures audio from an earbud ($d_1$), runs KWS~\cite{kws}, and sends results to a ring ($d_4$). Similarly, in Workload 4, pipeline 8 (object detector) captures images on glasses ($d_2$), runs MobileNetV2~\cite{sandler2018mobilenetv2} across devices, and sends results to a ring ($d_4$). For each pipeline, various execution plans are generated by leveraging the four AI accelerators, including splitting models at different layers and assigning chunks to different devices.

\begin{table}[t]
\centering
\caption{Workload and model description used in this study. Sizes are all in bytes.}
\label{tab:models}
%\vspace{-0.4cm}
\resizebox{\columnwidth}{!}{%
\begin{tabular}{ccrrrr}
\toprule
\multicolumn{1}{c}{\textbf{Workload}} & \multicolumn{1}{c}{\textbf{Pipeline}} & \multicolumn{1}{r}{\textbf{Models}} & \multicolumn{1}{r}{\textbf{Model Size}} & \multicolumn{1}{r}{\textbf{Input Size}} & \multicolumn{1}{r}{\textbf{Avg. Out Size}} \\ \hline
\multirow{3}{*}{\textbf{1}}           & 1                                     & ConvNet5                  & 71158                                   & 28$\times$28$\times$1                                 & 14031                                      \\
                                      & 2                                     & ResSimpleNet\cite{he2016deep}               & 381792                                  & 32$\times$32$\times$3                                 & 11217                                      \\
                                      & 3                                     & UNet\cite{ronneberger2015u}                       & 279084                                  & 48$\times$48$\times$48                                & 74547                                      \\ \hline
\multirow{3}{*}{\textbf{2}}           & 4                                     & KWS\cite{kws}                        & 169472                                  & 128$\times$128$\times$1                               & 7976                                       \\
                                      & 5                                     & SimpleNet\cite{hasanpour2016lets}                  & 166448                                  & 32$\times$32$\times$3                                 & 9237                                       \\
                                      & 6                                     & WideNet\cite{hasanpour2016lets}                    & 313700                                  & 32$\times$32$\times$3                                 & 10091                                      \\ \hline
\textbf{3}                            & 7                                     & EfficientNetV2\cite{tan2021efficientnetv2}             & 627220                                  & 32$\times$32$\times$3                                 & 66468                                      \\ \hline
\textbf{4}                            & 8                                     & MobileNetV2\cite{sandler2018mobilenetv2}                & 821164                                  & 32$\times$32$\times$3                                 & 296318                                    \\
\bottomrule
\end{tabular}
%\vspace{-0.4cm}
}
\end{table}

\begin{figure*}[t]
  \centering
  \includegraphics[width=0.99\linewidth]{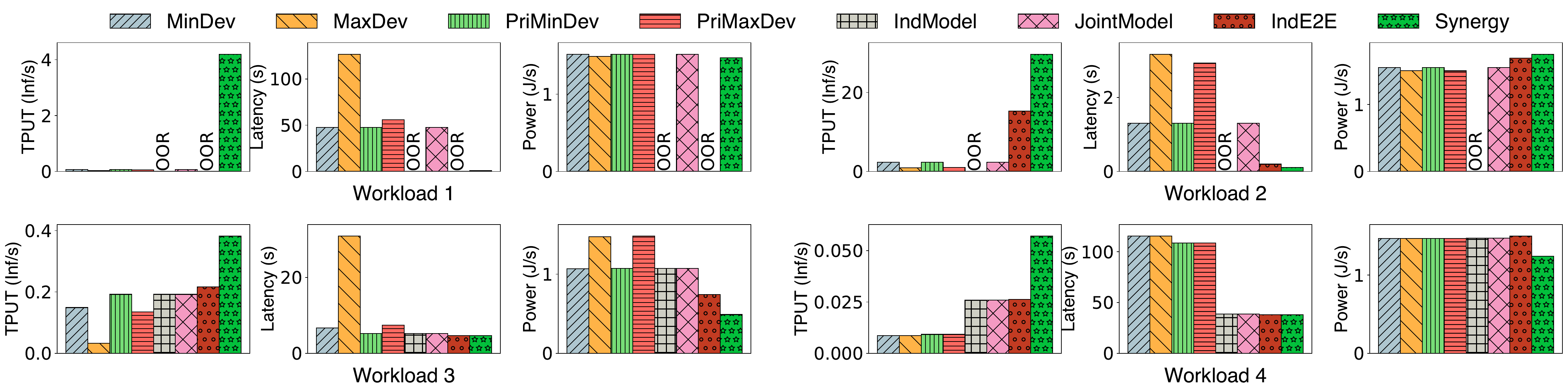}
%  \vspace{-0.5cm}
  \caption{Throughput, latency, and power consumption result for four different workloads.}
  \vspace{-0.3cm}
  \label{fig:overall_performance}
\end{figure*}

\subsubsection{Baselines} To the best of our knowledge, there are no existing studies that address model splitting in multi-pipeline scenarios across multiple devices. Therefore, we devised several baselines based on different rationales and adapted state-of-the-art splitting algorithms to evaluate their effectiveness in our environments. We consider 7 baselines. The first 4 are heuristic baselines that consider the resource usage of other pipelines when selecting the holistic collaboration plan. The last 3 are based on state-of-the-art algorithms. 

\textbf{MinDev}: This heuristic aims to avoid model splitting across devices as much as possible. The rationale is that using fewer devices would reduce communication overhead between devices, thereby increasing throughput. For each pipeline, it selects an execution plan that uses the minimum number of devices to run its model, considering the resource usage of previously selected plans, similar to \system{}.

\textbf{MaxDev}: In contrast, MaxDev focuses on maximizing model splitting over distributed AI accelerators, with the rationale that more devices could enhance task parallelization, thereby improving throughput. Unlike MinDev, MaxDev selects an execution plan that splits the model to all available devices.

\textbf{PriMinDev}: Prioritized MinDev (PriMinDev) enhances MinDev by prioritizing splitting points and device assignment order. For each pipeline, it selects an execution plan that minimizes intermediate output sizes from devices, while using the fewest possible devices. When selecting the device, it prioritizes MAX78002 over MAX78000 to reduce splitting.

\textbf{PriMaxDev}: PriMaxDev is the same as PriMinDev except that it considers execution plans that involve all devices.

\textbf{IndModel}: \rev{State-of-the-art model partitioning methods~\cite{kang2017neurosurgeon, dads2019dynamic, qdmp2020towards, laskaridis2020spinn, jeong2018ionn} primarily determine the optimal split execution plan based on metric estimations. However, these methods are designed for single-model optimization and do not directly extend to our multi-model scenario. We categorize them under IndModel and adapt them for our use case. Specifically, IndModel selects the best-split execution plan for each pipeline independently, without a holistic view. The final collaborative execution plan is then formed by aggregating these independently selected plans.}

% \textbf{IndModel}: Inspired by state-of-the-art model partitioning methods~\cite{kang2017neurosurgeon, dads2019dynamic, qdmp2020towards, laskaridis2020spinn, jeong2018ionn}, IndModel selects the best-split execution plan based on metric estimations, but without a holistic view. IndModel independently selects the optimal execution plan for each pipeline and forms the final holistic collaboration plan by aggregating the selected plans.

\textbf{JointModel:} IndModel may lead to out-of-resource (OOR) errors if the cumulative plan exceeds available resources. To prevent this, JointModel, a multi-tenant version of IndModel, conducts a joint resource assessment similar to \system{}. 

\textbf{IndE2E}: IndE2E  advances IndModel by incorporating source and target devices alongside model splitting. Each pipeline independently selects the execution plan expected to yield the highest throughput by considering end-to-end (E2E) latency, from sensing to output delivery. However, it does not account for the resource usage of other pipelines.

\subsubsection{Performance Metrics}
We consider three performance metrics: throughput, latency, and energy. Throughput is defined as the total number of model inferences per second, while latency is the time taken to execute an end-to-end holistic collaboration plan, respectively. For energy, we report the average power consumption for the execution of a holistic collaboration plan. We measured power consumption with Monsoon power monitor~\cite{Monsoon}. %with a sampling rate of 5 kHz.  

\subsection{Overall Performance}~\label{sec:evaluation:overall}
Figure~\ref{fig:overall_performance} shows the result of different methods for the four workloads. In terms of throughput, \system{} consistently outperforms all baselines owing to its holistic decision-making process and adaptive task parallelization. \system{} on average shows 23.0$\times$ higher throughput than the baselines. In Workload 1 and 2 where multiple pipelines run concurrently, IndModel (state-of-the-art model partitioning methods) results in OOR situations. In Workload 3 and 4, \system{} continues to exhibit 1.8$\times$ and 2.2$\times$ higher throughput than the second best (IndE2E) respectively. The results indicate two key points. First, \system{} boosts throughput without OOR failures when multiple pipelines exist (Workload 1 and 2). Second, \system{} supports large models through the collaboration of distributed AI accelerators (Workload 3 and 4).

For the other metrics, we first found that \system{} reduces latency significantly by 73.9\% on average, compared to the baselines. This is because maximizing throughput is tightly related to reduced latency; to maximize the throughput, the pipeline should be streamlined across multiple processors, which in turn reduces the end-to-end latency. Interestingly, we found that \system{} reduces power consumption by 15.8\% on average, despite its significant throughput improvement of 23.0$\times$ that naturally entails more use of computation units. This is because the major source of power consumption is data transmission between devices, and maximizing throughput in \system{} naturally minimizes data communication cost. 
% \begin{table}[t]
% \centering
%   \caption{Ablation study of \system{}.}\label{tab:ablation}
% %  \vspace{-0.4cm}
% \resizebox{0.8\columnwidth}{!}{%
% \begin{tabular}{cccccc}
% \toprule
% JRC & STT & PSR & ATP & \multicolumn{1}{c}{TPUT (Wrkld 1)} & \multicolumn{1}{c}{TPUT (Wrkld 2)} \\ \hline
%  &  &  &  & OOR & OOR \\\hline
% \checkmark &  &  &  & 0.06 & 2.30 \\\hline
% \checkmark & \checkmark &  &  & 0.92 & 15.28 \\\hline
% \checkmark & \checkmark & \checkmark &  & 2.72 & 15.28 \\\hline
% \checkmark & \checkmark & \checkmark & \checkmark & \textbf{4.20} & \textbf{29.67}\\\bottomrule

% \end{tabular}%
% }
% \vspace{-.4cm}
% \end{table}% Please add the following required packages to your document preamble:
% \usepackage{multirow}

\begin{table*}[]
\centering
  \caption{\rev{Ablation study of \system{}.}}\label{tab:ablation}
% \color{blue} % Apply blue color to the entire table
\begin{tabular}{cccccrrcrr}
\toprule
\multirow{2}{*}{JRC} & \multirow{2}{*}{STT} & \multirow{2}{*}{PSR} & \multirow{2}{*}{ATP} & \multicolumn{3}{c}{Workload 1} & \multicolumn{3}{c}{Workload 2} \\ \cline{5-10} 
 &  &  &  & \multicolumn{1}{l}{TPUT (inf/s)} & \multicolumn{1}{l}{Latency (s)} & \multicolumn{1}{l}{Power (J/s)} & \multicolumn{1}{l}{TPUT (inf/s)} & \multicolumn{1}{l}{Latency (s)} & \multicolumn{1}{l}{Power (J/s)} \\ \hline
 &  &  &  & OOR & OOR & OOR & OOR & OOR & OOR \\ \hline
\checkmark &  &  &  & 0.06 & 47.63 & 1.52 & 2.30 & 1.31 & \textbf{1.55} \\ \hline
\checkmark & \checkmark &  &  & 0.92 & 3.25 & 1.59 & 15.28 & 0.20 & 1.69 \\ \hline
\checkmark & \checkmark & \checkmark &  & 2.72 & 1.10 & 1.61 & 15.28 & 0.20 & 1.69 \\ \hline
\checkmark & \checkmark & \checkmark & \checkmark & \textbf{4.20} & \textbf{0.86} & \textbf{1.47} & \textbf{29.67} & \textbf{0.10} & 1.75
\\\bottomrule
\end{tabular}
\end{table*}

\subsubsection{Ablation Study}

We perform an ablative study to understand the effectiveness of each technical component of \system{}. Table~\ref{tab:ablation} shows the result. JRC refers to the joint resource consideration among pipelines, and STT refers to device mapping accounting for source and target tasks (\S\ref{sec:runtime:holistic}). PSR means progressive search space reduction (\S\ref{sec:runtime:planning}), and ATP means adaptive task parallelization (\S\ref{sec:runtime:deployment}). If none of these are applied, it is the same as the state-of-the-art approach (IndModel). If all are checked, it becomes \system{}. \rev{Overall, the throughput and latency improve as each of our components is added. Improving IndModel by addressing resource conflicts among pipelines to avoid OOR (JRC) is the same as JointModel in our baselines. Note that merely resolving the resource conflicts is still far from \system{}'s performance. Incorporating other technical components, STT, PSR, and ATP shows 6.9, 7.6, and 14.3$\times$ higher throughput compared to JRC-only, respectively. Similarly, the latency was reduced by 93.0\%, 97.3\%, and 98.0\% compared to JRC-only, owing to the parallelization of the pipelines. We also noted that utilizing all components increased power consumption by around 12.9\% compared to JRC-only, due to the increased use of computation units. These findings are aligned with the earlier analysis of the overall performance.}

%Overall, the throughput improves as each of our components is added. Improving IndModel by addressing resource conflicts among pipelines to avoid OOR is the same as JointModel in our baselines. Note that merely resolving the resource conflicts is still far from \system{}. Incorporating other technical components, STT, PSR, and ATP shows 6.9, 7.6, and 14.3$\times$ higher throughput compared to JRC-only, respectively. 
\subsection{In-Depth Analysis}

\subsubsection{Runtime Environment Changes}

We investigate how \system{} adapts to changes in the runtime environment. %We consider two scenarios: variations in the number of devices and the number of pipelines.

\begin{figure}[t]
  \centering
  \includegraphics[width=0.9\linewidth]{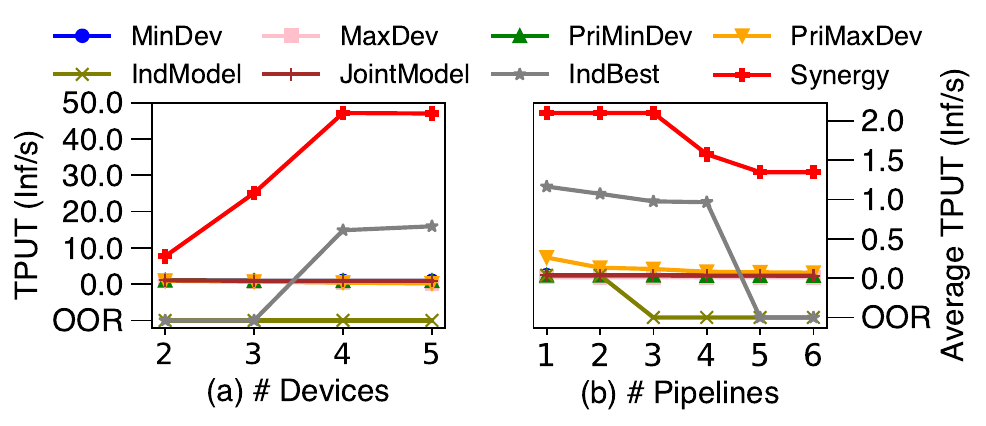}
%  \vspace{-0.5cm}
  \caption{Impact of runtime environment changes: (a) number of devices and (b) number of pipelines. 
  }
  \vspace{-0.5cm}
  \label{fig:num_dev_mksr}
\end{figure}

\textbf{Number of devices:} 
In this experiment, we explore \system{}'s scalability to the change of the number of MAX78000 devices. We increase the number of devices from two to five while running the same set of four pipelines with ConvNet5, KWS, SimpleNet, and ResSimpleNet. Figure~\ref{fig:num_dev_mksr} (a) shows the result.
Interestingly, \system{} significantly outperforms the baselines as the number of devices increases. This is due to \system{}'s strategic consideration of accelerator assignments, effectively minimizing communication overhead between source and target devices. Conversely, except for IndE2E, the throughput gains for all other baselines are not notable, even with additional devices. 
Another interesting observation is that using more devices does not always lead to higher throughput. In the case of \system{}, the throughput saturates once the number of devices reaches 4. This is because, beyond optimal distribution, further splitting of models fails to contribute to additional throughput gains.

\textbf{Number of pipelines:} We assessed \system{}’s performance with a varying number of pipelines, incrementally increasing from one to six: UNet, ConvNet5, SimpleNet, KWS, ResSimpleNet, and WideNet, using four MAX78000s. To understand resource competition, we report the \emph{average} throughput across pipelines, i.e., the ratio of completed pipelines per second to the total number of pipelines. Figure~\ref{fig:num_dev_mksr} (b) shows a downward trend in average throughput as the number of pipelines increases for all baselines due to competition for resources. Nonetheless, \system{} consistently outperforms the baselines, achieving an average throughput of 1.35 with six pipelines, 19.4$\times$ higher than the second best (PriMaxDev).

\subsubsection{Composing Heterogeneous Accelerators}

\begin{figure}[t]
  \centering
  \includegraphics[width=0.8\linewidth]{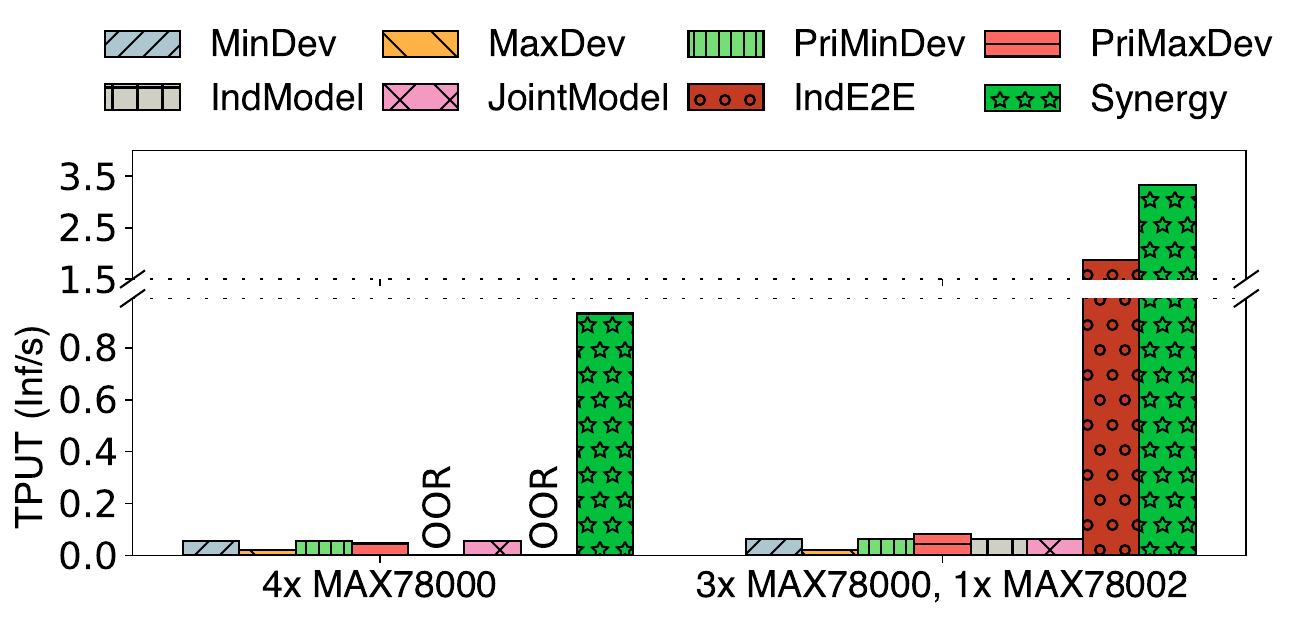}
%  \vspace{-0.15in}
  \caption{Effect of accelerator composition.}
  \vspace{-0.2in}
  \label{fig:dev_type}
\end{figure}

To examine \system{}'s effectiveness in heterogeneous accelerator resources, we conducted an experiment where one of four MAX78000 devices was substituted with more resource-capable MAX78002. The workload is comprised of three pipelines: ConvNet5, UNet, and EfficientNet. Figure~\ref{fig:dev_type} shows the results in two different setups.
The inclusion of a higher-resource device generally led to an improvement in throughput.  
With four MAX78000s, \system{} achieves a total throughput of 0.93, outperforming the second best (JointModel) by a factor of 16.4. This throughput further increases to 3.33 when incorporating one MAX78002. PriMinDev, which allocates all models exclusively to the single MAX78002, results in a significantly low throughput of 0.06. \rev{This discrepancy demonstrates why simply offloading model execution to a more powerful device (MAX78002) is not always the best approach. It also underscores the need to account for communication overhead between source and target devices. Additionally, we observed that while IndE2E results in an OOR condition with four MAX78000s, it achieves the second-highest throughput when using the MAX78002. This suggests that IndE2E performs well in resource-rich environments but fails to manage resources carefully in resource-constrained settings.}
%This discrepancy highlights the importance of consideration for the communication overhead associated with source and target devices.

\begin{figure}[t]
  \centering
  \includegraphics[width=0.77\linewidth]{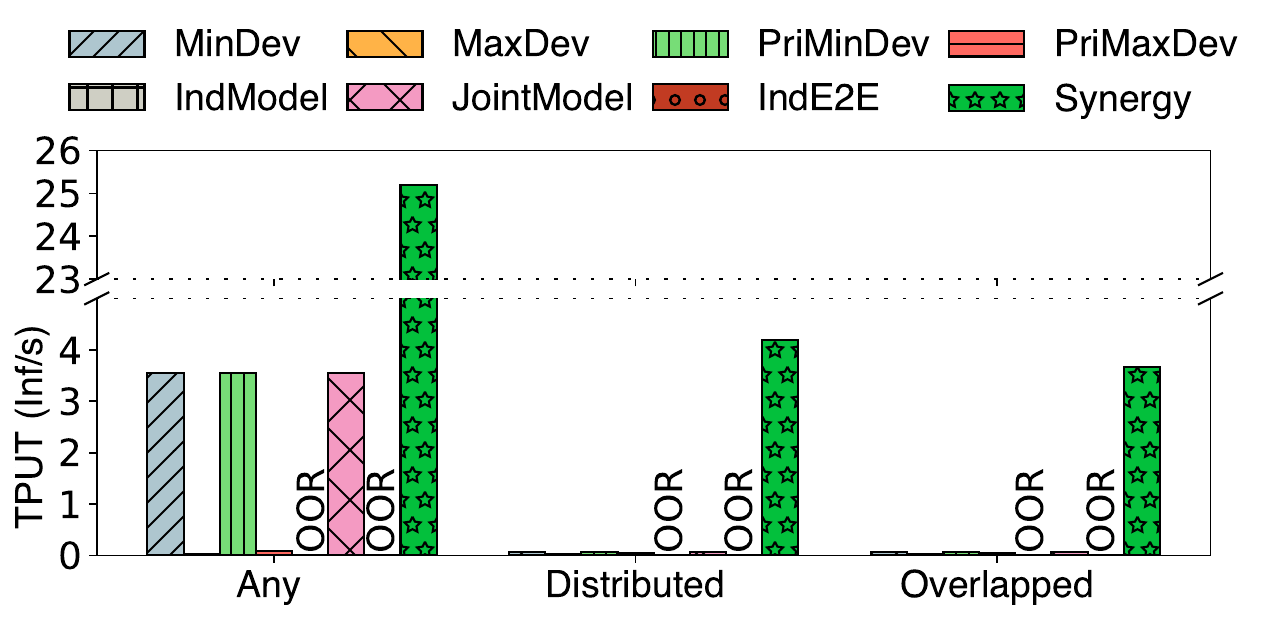}
%  \vspace{-0.20in}
  \caption{Effect of source and target mappings.}
%  \vspace{-0.15in}
  \label{fig:srcdst}
\end{figure}

\subsubsection{Source and Target Mapping}

We assessed source and target mapping effects in three scenarios: (i) any device can be source or target (Any), (ii) source/target devices are evenly allocated (Distributed), and (iii) the same device is used as both source and target across pipelines (Overlapped). The Distributed setup matches Workload 1 from \S\ref{sec:evaluation:overall}, while the Overlapped and Any scenarios only differ in their source and target device mappings. As shown in Figure~\ref{fig:srcdst}, the Overlapped scenario has the lowest throughput due to communication bottlenecks, while the Any scenario has the highest throughput by distributing communication costs. The result indicates \system{}’s robustness in execution planning, considering both source and target factors.

\begin{table}[t]
\centering
\caption{Comparison among different objectives.}
\label{tab:policy}
%\vspace{-0.4cm}
\resizebox{0.75\columnwidth}{!}{%
\begin{tabular}{lrrr}
\toprule

\multicolumn{1}{c}{Workload 1} & \multicolumn{1}{c}{TPUT (inf/s)} & \multicolumn{1}{c}{Latency (s)} & \multicolumn{1}{c}{Power (J/s)}  \\ \hline
TPUT-max & \textbf{4.20} & \textbf{0.86} & 1.47 \\ \hline
Latency-min & 3.15 & \textbf{0.86} & 1.42 \\ \hline
Power-min & 0.19 & 27.17 & \textbf{1.22} \\ 
\bottomrule 
\noalign{\smallskip} % Adds a small vertical space
\noalign{\smallskip} % Adds a small vertical space
\toprule
\multicolumn{1}{c}{Workload 2} & \multicolumn{1}{c}{TPUT (inf/s)} & \multicolumn{1}{c}{Latency (s)} & \multicolumn{1}{c}{Power (J/s)}  \\ \hline
TPUT-max & \textbf{29.67} & \textbf{0.10} & 1.37 \\ \hline
Latency-min & \textbf{29.67} & \textbf{0.10} & 3.21 \\ \hline
Power-min & 1.37 & 3.21 & \textbf{1.06} \\ \bottomrule
\end{tabular}%
}
\vspace{-.3cm}
\end{table}

\subsubsection{Different Objectives}\label{sec:evaluation:target_metric}
While \system{} defaults to maximizing system-wide throughput (TPUT-max), it can also minimize latency (Latency-min) or power consumption (Power-min). Table~\ref{tab:policy} demonstrates \system{}'s ability to achieve these different objectives effectively. Each objective excels in its metric, though identical latencies do not ensure the same holistic collaboration plan. Notably, maximizing throughput provides a balanced performance with low latency and near-minimal power use. For example, in workload 1, TPUT-max achieves 22.1$\times$ higher throughput with only 1.2$\times$ more power consumption compared to Power-min. This is because maximizing throughput reduces latency and communication overhead, the primary power consumption source.

\begin{figure}[t]
  \centering
  \includegraphics[width=0.77\linewidth]{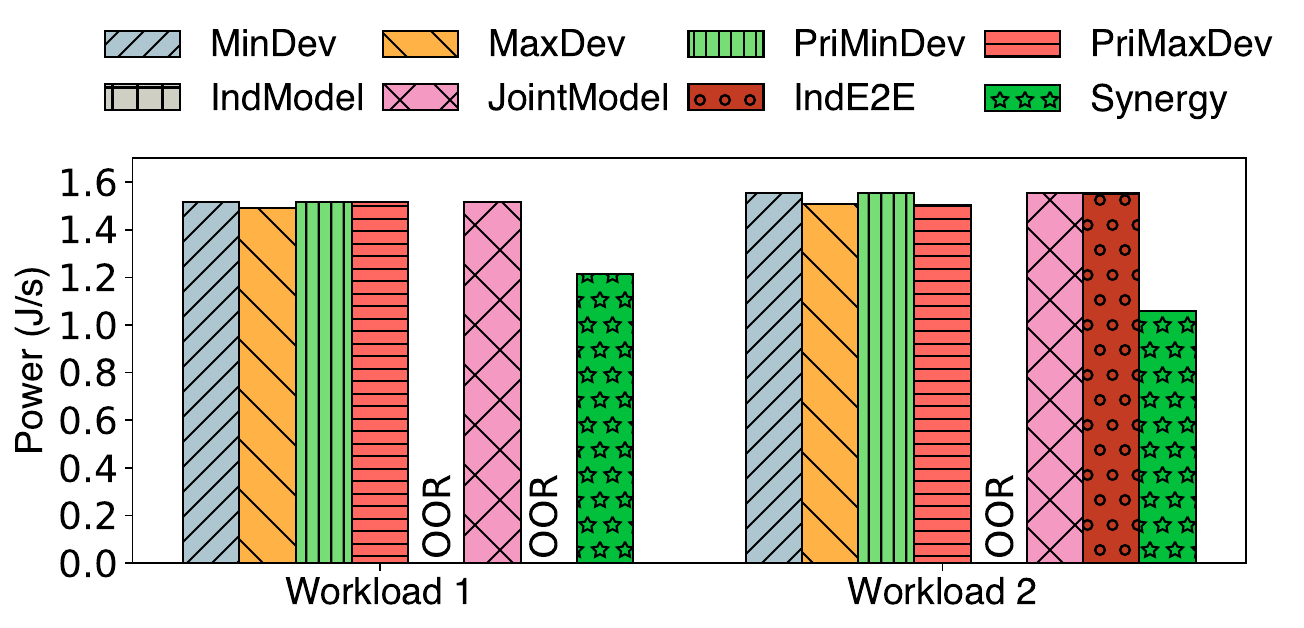}
%  \vspace{-0.20in}
  \caption{\rev{Effect of the objective: minimizing power consumption.}}
%  \vspace{-0.15in}
  \label{fig:power_objective}
\end{figure}

\rev{
We further investigate the effectiveness of \system{}'s orchestration by taking Power-min as an example. Figure~\ref{fig:power_objective} shows the average power consumption for Workload 1 and Workload 2 when Power-min is set as an objective. The baselines also select the execution plan, prioritizing the minimization of power consumption. The results show that \system{} executes multiple workloads while minimizing power consumption and avoiding OOR.
}

\section{Related work}
\rev{
% \noindent
\textbf{TinyML:} TinyML represents a research field in machine learning techniques, aiming to bring AI capabilities to the most resource-constrained devices, such as MCUs. These devices typically have tens to hundreds of kilobytes of SRAM, and most of the research efforts in this domain have focused on minimizing model size. Existing studies have explored mainly three techniques: model pruning~\cite{liberis2023differentiable}, model quantization~\cite{rusci2020memory}, and neural architecture search (NAS)~\cite{liberis2021munas, fedorov2019sparse}. 
Model pruning~\cite{liberis2023differentiable} reduces model complexity by identifying and eliminating unnecessary parameters that contribute minimally to output accuracy. Model quantization~\cite{rusci2020memory} decreases both model size and computational intensity of operations by reducing the precision of the numerical values used to represent a model's parameters. NAS~\cite{liberis2021munas, fedorov2019sparse} is an automated method for discovering the best neural network architectures that are tailored to the resource constraints of a specific device, while balancing efficiency and accuracy.
In contrast, our work focuses on supporting multiple or large AI models without compromising accuracy, by dynamically composing distributed AI accelerators. However, the combined power of tiny AI accelerators may still be insufficient to support unmodified, off-the-shelf large AI models. We envision that \system{} can benefit from these TinyML techniques to accommodate larger AI models.
}

% \noindent
\textbf{Model partitioning:} While there are very few attempts for partitioning AI models over multiple MCU-equipped devices, there have been active research efforts for layer-wise model partitioning over resource-limited embedded and mobile devices. In common, these methods~\cite{kang2017neurosurgeon, dads2019dynamic, qdmp2020towards, laskaridis2020spinn, jeong2018ionn, cas2021context, banitalebi2021autosplit} allocates few initial layers of a DNN on mobile/embedded device and the latter in edge or cloud server. Intermediate output from the initial layer execution is transmitted to powerful resources such as the cloud or nearby edge device where the subsequent part of the model is executed. They adapt the splitting layer depending on network status and server load. While \system{} also adopts vertical partitioning for distributed inferences, existing model partitioning studies lack the (i) end-to-end perspective of the on-body AI app execution (sensor and interface operations as well as model execution) and (ii) consideration of concurrent execution of multi-tenant models. Our experiments demonstrate that \system{} outperforms these studies by solving the challenges.

% \noindent
\textbf{DNN workload distribution:} 
An application pipeline often consists of a (conditional) sequence of multiple DNN models. Several distributed systems have been proposed to balance model workloads across distributed devices~\cite{zeng2020distream, jang2021microservice, dong2023watchdog}. 
Their key difference with model partitioning methods is to treat a model as a primitive execution unit and focus more on scheduling \textit{model} execution over distributed devices (without model splitting). For example, a face classification task can be executed on a different device following object detection on the initial device, thereby leveraging distributed resources from multiple devices~\cite{zeng2020distream, jang2021microservice}. \system{} shares the same high-level objective of distributing multiple model workloads into multiple devices. However, our target environment is on-body devices with AI accelerators, which brings resource challenges. To overcome the limited resources of these devices, we adopt model partitioning on top of workload distribution and devise a tailored solution for tiny AI accelerators.

% \noindent
\textbf{Model serving systems:} Several platforms, such as TensorFlow serving~\cite{tensorflowserve}, Sagemaker~\cite{sagemaker}, and Azure ML~\cite{azureml}, have been proposed to facilitate model inference serving by offering containerized environments for model execution in diverse devices. Research platforms, such as Velox~\cite{crankshaw2014velox} and Clipper~\cite{crankshaw2017clipper}, focus on low-latency prediction serving, together with optimizing cloud server performance. However, in wearable computing, the main challenge lies in dynamically composing distributed tiny AI accelerators. Our \emph{device-agnostic programming interface} addresses this by mapping software logic to physical resources and optimizing performance by considering pipeline interdependencies and resource usage.

\rev{
\textbf{Middleware for body sensor networks:} Several decades ago, body sensor networks (BSNs) received attention for recognizing human contexts using sensory signals from multiple wearable devices. Early work focused on fusion techniques to effectively and robustly concatenate multiple data streams~\cite{ordonez2016deep,peng2018aroma,yao2017deepsense,yao2018qualitydeepsense,vaizman2018context} when different devices introduced several challenges, e.g., time synchronization, missing data, and varying sampling rates. Another focus was handling dynamic device availability~\cite{zappi2008activity,kang2008seemon,kang2010orchestrator,lee2013active,keally2011pbn}, i.e., when the set of available devices changes over time. To tackle this problem, many studies have explored the effects of different characteristics on recognition accuracy, e.g., types, compositions, and placements of sensors, and have proposed methods to dynamically select the best sensor based on predefined parameters. 

Although these works provide a foundation for sensing and processing in multi-device environments, their consideration of sensing pipelines and target devices has been limited to support on-body AI apps we envision. First, Sensing pipelines in BSNs was simple, so pipeline partitioning across devices was not actively studied. Second, due to insufficient processing capabilities, wearable devices were primarily treated as sensor data streamers, with most processing assumed to be run on a smartphone. However, we argue that on-body sensing pipelines have evolved with the introduction of numerous DNN-based AI models, and wearable devices have begun to feature on-device AI capabilities, with the emergence of tiny AI accelerators. This insight necessitates revisiting existing strategies for on-body sensing and processing in multi-wearable environments. To the best of our knowledge, our work is the first to efficiently support on-body AI applications on wearable devices equipped with tiny AI accelerators.
}
\section{Conclusion}
We presented \system{}, a novel system that supports on-body AI apps through the collaboration of tiny AI accelerators on wearables. \system{}’s device-agnostic programming interface simplifies integrating diverse AI applications. Its runtime dynamically distributes model execution tasks to available device resources for optimal inference. Our evaluation showed \system{} achieves higher throughput than baselines.

Future work aims to enhance collaboration among AI accelerators by expanding from layer-wise model splitting to include channel-wise splitting~\cite{mao2017modnn, mao2017mednn, zeng2020coedge, zhao2018deepthings, hou2022distredge, edgeflow2022distributed}, enabling \system{} to handle larger inputs and optimize performance through a combination of both techniques~\cite{pang2023adamec}. Additionally, we plan to extend \system{} to support more complex model architectures, such as sensor fusion and conditional chaining of multiple models, to meet real-world application demands. Lastly, we aim to enrich the device-agnostic programming interface by adding features like detailed taxonomy for sensor and target device mapping and supporting customized functions, thereby providing developers with greater flexibility in designing on-body AI apps.

\bibliographystyle{IEEEtran}
\bibliography{reference.bib}

\end{document}